\documentclass[conference]{IEEEtran}
\IEEEoverridecommandlockouts
\usepackage{amsmath,amssymb,amsfonts,amsthm}
\usepackage{textcomp}
\usepackage{xcolor}
\usepackage{cite}
\usepackage{bbm}
\usepackage{bm}
\usepackage[hidelinks]{hyperref}
\usepackage{graphicx}

\usepackage{algorithm}
\usepackage{algpseudocode} % Or 
\newtheorem{theorem}{Theorem}
\newtheorem{corollary}[theorem]{Corollary}   % Shares counter with theorem
\newtheorem{remark}{Remark}

\DeclareMathOperator*{\argmin}{arg\,min}

\begin{document}

\title{Information Gradient for Directed Acyclic Graphs: A Score-based Framework 
for End-to-End Mutual Information Maximization}

\author{
\IEEEauthorblockN{Tadashi Wadayama}
\IEEEauthorblockA{Nagoya Institute of Technology\\
\texttt{wadayama@nitech.ac.jp}}
}

\maketitle

\begin{abstract} 
  This paper presents a general framework for end-to-end mutual information maximization 
  in communication and sensing systems represented by stochastic directed acyclic graphs (DAGs). We derive a 
  unified formula for the (mutual) information gradient with respect to arbitrary internal 
  parameters, utilizing marginal and conditional score functions. We demonstrate 
  that this gradient can be efficiently computed using vector-Jacobian products (VJP) 
  within standard automatic differentiation frameworks, enabling the optimization 
  of complex networks under global resource constraints. Numerical experiments on 
  both linear multipath DAGs and nonlinear channels validate the proposed framework; 
  the results confirm that the estimator, utilizing score functions learned 
  via denoising score matching, accurately reproduces ground-truth gradients 
  and successfully maximizes end-to-end mutual information.
  Beyond maximization, we extend our score-based framework to a novel unsupervised 
  paradigm: digital twin calibration via Fisher divergence minimization.
\end{abstract}


\section{Introduction} 
\label{sec:introduction}

Designing complex communication and sensing systems to maximize information 
flow is a central challenge in modern engineering. From next-generation 
wireless networks aiming for semantic communication \cite{Gunduz2023} to distributed sensor fusion 
in IoT, the paradigm is shifting from optimizing individual modules to \textit{end-to-end} system design. 
In these scenarios, mutual information (MI) \cite{cover2006} serves as the gold standard objective function, 
capturing nonlinear dependencies that simple correlations miss. However, maximizing MI in such high-dimensional, multi-stage systems presents 
a formidable computational barrier. Unlike deterministic loss functions (e.g., mean squared error) where gradients are 
easily computed via backpropagation, 
MI depends on the joint probability density, which evolves in complex ways through the system dynamics. Existing variational bounds focus 
primarily on \textit{estimating} the MI value; utilizing them 
for gradient-based optimization of internal system parameters 
often leads to high variance and instability. 
Essentially, we lack a scalable mechanism to "backpropagate" information measures through complex 
stochastic graphs.

In a companion paper \cite{wadayama2025-b}, we took a first step toward addressing this challenge 
by introducing the \textit{information gradient} framework. This theory establishes an exact identity linking 
the gradient of MI to \textit{score functions} \cite{cover2006, hyvarinen2005}—the gradients of 
log-likelihoods. By leveraging modern score matching 
techniques \cite{vincent2011, song2019, song2021}, 
this framework allows for efficient gradient estimation 
without closed-form likelihoods. However, the scope 
of \cite{wadayama2025-b} was limited to single-stage 
channels of the form $Y=f_{\bm \eta}(X) + Z$.
Real-world systems are rarely this simple; 
they involve intricate topologies with branching paths, 
merging signals, and multi-hop processing, 
effectively forming stochastic networks. 
Applying the basic formula to such architectures 
is non-trivial, as a single parameter can influence 
the final output through multiple interacting pathways, 
complicating the derivation of the score function and 
the associated chain rules.

This paper generalizes the information gradient framework 
to arbitrary stochastic systems represented 
by \textit{directed acyclic graphs (DAGs)}. We view the 
system as a computational graph where nodes represent stochastic variables 
and edges represent parameterized nonlineartransformations. 
This means that any stochastic network system implementable 
in modern automatic differentiation frameworks 
(e.g., PyTorch, TensorFlow, JAX) can now be optimized 
for MI using our score-based estimator.

Our main contribution is a unified gradient formula 
applicable to any internal parameter within a DAG, 
regardless of its depth or topological complexity. 
Crucially, we bridge the gap between information theory 
and deep learning implementation: we show that 
the generalized information gradient can be efficiently 
computed using standard \textit{vector-Jacobian products (VJP)} \cite{Baydin2018}. 
Moreover, we demonstrate that canonical multi-user models, 
such as multiple access channels (MAC) and broadcast channels (BC), are naturally 
represented as DAGs. Consequently, our method serves as a unified numerical solver 
for exploring the achievable rate regions of complex nonlinear channels 
where closed-form analytical characterizations are intractable. To rigorously validate 
this capability, we show that our score-based information gradient exactly 
reproduces the known capacity region of the Gaussian MAC.

This generalization opens up broad practical applications. 
It enables the end-to-end optimization of transmitter and 
receiver neural networks over unknown nonlinear channels, 
a key requirement for {\em semantic communications}. Furthermore, 
we extend the framework to a novel unsupervised paradigm: 
\textit{digital twin calibration}. By minimizing the Fisher 
divergence—a natural byproduct of our score-based formulation—
we propose a method to align simulation parameters 
with physical reality using only output observations.

\section{Preliminaries}
\label{sec:preliminaries}

In this section, we introduce the basic notations 
and review key concepts regarding the fundamental 
information gradient identity 
established in our companion paper \cite{wadayama2025-b}.

\subsection{Notations}
We use boldface lowercase letters (e.g., $\bm{x}$) to denote vectors and boldface uppercase letters (e.g., $\bm{A}$) for matrices. Random vectors are denoted by uppercase letters (e.g., $X, Y$), and their realizations by lowercase letters (e.g., $\bm{x}, \bm{y}$).
The probability density function (PDF) of a continuous random vector $X$ is denoted by $p_X(\bm{x})$.
The expectation operator is denoted by $\mathbb{E}[\cdot]$.
We use $h(X) \equiv -\mathbb{E}[\log p_X(X)]$ to denote 
the differential entropy of $X$, and $I(X; Y) \equiv h(Y) - h(Y|X)$ for 
the mutual information between $X$ and $Y$ \cite{cover2006}.
The Jacobian matrix of a vector-valued differentiable function 
$\bm{f}(\bm{x})$ with respect to $\bm{x}$ is denoted by $D_{\bm{x}}\bm{f} 
\equiv \frac{\partial \bm{f}}{\partial \bm{x}}$.

\subsection{Score Functions}
The gradient of the log-likelihood with respect to the data variable is termed 
the \textit{score function} \cite{cover2006}.
For a random vector $Y \in \mathbb{R}^d$ with PDF $p_Y(\bm{y})$, 
the (marginal) score function is defined as
\begin{equation}
\bm{s}_Y(\bm{y}) \equiv \nabla_{\bm{y}} \log p_Y(\bm{y}).
\end{equation}
Similarly, the conditional score function for $Y$ given $X$ is 
$\bm{s}_{Y|X}(\bm{y}|\bm{x}) \equiv \nabla_{\bm{y}} \log p_{Y|X}(\bm{y}|\bm{x})$.
A fundamental property of the score function is that its expectation under the corresponding distribution is zero, provided that the PDF vanishes at the boundary:
\begin{align}
\mathbb{E} [\bm{s}_Y(Y)] 
    &= \int p_Y(\bm{y}) \nabla_{\bm{y}} \log p_Y(\bm{y}) d\bm{y} \notag \\
    &= \int \nabla_{\bm{y}} p_Y(\bm{y}) d\bm{y} = \bm{0}.
\label{eq:score_zero_mean}
\end{align}
This property plays a crucial role in simplifying gradient derivations.

\subsection{Related works}

Our framework including \cite{wadayama2025} is grounded in the deep correspondence between information theory and thermodynamics, 
specifically the interpretation of noise injection as a diffusion process. 
This perspective originates from Stam \cite{stam1959}, who utilized the de Bruijn identity—relating 
the derivative of entropy to Fisher information—to establish the entropy power inequality. 
Building on this diffusion-based view, Brown \cite{brown1982} and Barron \cite{barron1986entropy} 
developed the entropic central limit theorem, proving 
the monotonic convergence of Fisher information. 
In the context of Gaussian channels, Guo, Shamai, and Verdú \cite{guo2005} unified these concepts 
through the I-MMSE relationship, identifying the derivative of mutual information with the minimum mean-square error.

While these classical studies primarily employed these relationships for theoretical proofs of inequalities, 
our work repurposes them for practical computation. 
We extend the utility of these fundamental identities from theoretical analysis to data-driven optimization. 
By generalizing these connections to arbitrary stochastic DAGs and estimating the 
requisite score functions from data, we transform these classical thermodynamic links 
into a scalable engine for end-to-end mutual information maximization.

\subsection{Information Gradient Identity}
In our companion paper \cite{wadayama2025-b}, we derived a fundamental identity that relates the gradient of mutual information to score functions.
Consider a general parameterized system where the output is given by $Y = f_{\bm \eta}(X) + Z$, with input $X \in \mathbb{R}^{n}$, 
where $f_{\bm \eta}: \mathbb{R}^{n} \to \mathbb{R}^{m}$ is a differentiable 
function parameterized by $\bm{\eta} \in \mathbb{R}^d$,
and $Z \in \mathbb{R}^{m}$ is an additive noise vector.
Under standard regularity conditions, the gradient of the mutual information 
with respect to $\bm{\eta}$ can be expressed as
\begin{equation}
\nabla_{\bm{\eta}} I(X; Y) = \mathbb{E}\left[ (D_{\bm{\eta}} Y)^\top (\bm{s}_{Y|X}(Y|X) - \bm{s}_{Y}(Y)) \right].
\label{eq:general_info_grad}
\end{equation}
This identity states that the information gradient is the expected inner product of the Jacobian of the output with respect to the parameters, and a ``score difference'' vector.
While the paper \cite{wadayama2025-b} focuses on the theoretical derivation and 
properties of this identity for basic channel models, 
the present paper generalizes this formula 
for complex stochastic networks represented by general DAGs.

\subsection{Denoising Score Matching}

Since the true data distributions are typically unknown in complex systems, 
we must estimate the score functions from data.
Denoising score matching (DSM) \cite{vincent2011, song2019, song2021} is 
a highly effective technique for this purpose.
When the data $\bm{y}$ is generated by adding Gaussian noise 
$\bm{z} \sim \mathcal{N}(\bm{0}, \sigma^2 \bm{I})$ 
to a clean sample $\tilde{\bm{y}}$, 
i.e., $\bm{y} = \tilde{\bm{y}} + \bm{z}$, 
a score model $s_{\bm{\phi}}(\bm{y})$ 
parameterized by a parameter vector $\bm{\phi}$ can be trained 
by minimizing the following objective:
\begin{equation}
\mathcal{J}_{\text{DSM}}(\bm{\phi}) 
= \mathbb{E}_{\tilde{\bm{y}}, \bm{z}} 
\left[ \left\| \bm{s}_{\bm{\phi}}(\tilde{\bm{y}} + \bm{z}) 
+ \frac{\bm{z}}{\sigma^2} \right\|^2 \right].
\end{equation}
Minimizing this objective ensures that $\bm{s}_{\bm{\phi}}(\bm{y})$ approximates the true score $\nabla_{\bm{y}} \log p_Y(\bm{y})$.
We utilize this technique to learn the necessary score functions for gradient estimation.

While DSM are mainly used throughout this paper, 
it should be noted that the velocity field framework 
(e.g., flow matching \cite{lipman2023flowmatching} 
and rectified flow \cite{liu2023rectifiedflow}) 
could be employed to derive a score function in a data-driven manner.
Since there exists a one-to-one algebraic correspondence 
between the score function and the velocity field,
we can derive a score function from a velocity field.
In practice, training a velocity field might be 
significantly more stable than 
its score-based counterpart, especially in low-noise regimes. 
In such cases, the use of the velocity field framework 
for score function estimation might be more advantageous.

\section{Information Gradient for Cascaded Channels}

It is beneficial to consider the cascade case 
as a building block for more complex systems represented by DAGs.
In this section, 
we derive the information gradient for a cascade of 
deterministic functions with additive Gaussian noises.

\subsection{Problem setup}

Without loss of generality, we consider an $N$-stage cascaded 
nonlinear Gaussian channels.
 Let $X \in \mathbb{R}^{d_0}$ be the random input source vector following 
a prior probability density function (PDF) $p_X$. 
The cascaded channel consists of a sequence of $N$ computational blocks, 
where the output of the $i$-th block, denoted by $Y_i \in \mathbb{R}^{d_i}$, 
serves as the input to the $(i+1)$-th block.

\begin{figure}[htbp]
  \centering
  \includegraphics[width=\linewidth]{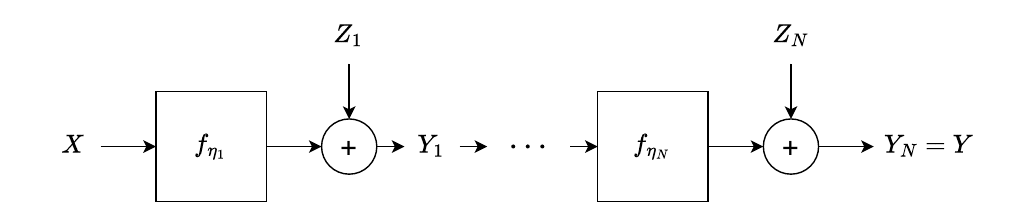}
  \caption{A schematic representation of an $N$-stage cascaded nonlinear Gaussian channel. Each stage $i$ consists of a deterministic mapping $f_{\bm \eta_i}$ followed by additive Gaussian noise $Z_i$. The system forms a Markov chain $X \to Y_1 \to \dots \to Y_N = Y$.}
  \label{fig:cascaded_channel}
\end{figure}

We assume that each block $i \in \{1, \dots, N\}$ consists of 
a parameterized deterministic mapping followed by 
an additive Gaussian noise channel. 
Specifically, the relationship between consecutive intermediate variables 
is given by
\begin{equation}
Y_i = f_{\bm \eta_i}(Y_{i-1}) + Z_i, \quad i=1, \dots, N,
\end{equation}
where $Y_0 \triangleq X$ is the system input, 
and $Y \equiv Y_N$ represents the final output of 
the cascaded channel. 
Here, $f_{\bm \eta_i}: \mathbb{R}^{d_{i-1}} \to \mathbb{R}^{d_i}$ is 
a deterministic differentiable function parameterized 
by a parameter vector $\bm \eta_i$. 
The noise vector $Z_i \in \mathbb{R}^{d_i}$ is 
assumed to be independent and identically distributed (i.i.d.)  Gaussian 
random variables with zero mean and covariance matrix 
$\sigma_i^2 \bm I\ (\sigma_i > 0)$,
which are independent of $X$ and $\{Z_{j}\}_{j \neq i}$.
Figure \ref{fig:cascaded_channel} shows 
a schematic representation of the cascaded channel.

Due to the noise injection at each stage, the mapping from $X$ 
to any intermediate output $Y_i$ (including the final output $Y = Y_N$) 
is stochastic. The entire cascade forms a Markov chain:
\begin{equation}
X \to Y_1 \to Y_2 \to \dots \to Y_N = Y.
\end{equation}
We assume that $P_Y$ and $P_{Y|X}$ exist and are differentiable.

Our primary goal is to maximize the end-to-end MI
between the initial input $X$ and the final output $Y$ 
by optimizing the set of parameters $\{\bm \eta_1, \dots, \bm \eta_N\}$. 
The optimization problem is defined by
\begin{equation}
\text{maximize}_{\{\bm \eta_i\}} \quad I(X; Y).
\end{equation}
To employ gradient-based optimization methods, 
we need to compute the gradient of this objective with respect 
to any intermediate parameter $\bm \eta_k$ ($1 \le k \le N$), 
i.e., $\nabla_{\bm \eta_k} I(X; Y)$.

\subsection{Derivation of the Gradient Formula}

We now derive the gradient of the mutual information 
$I(X; Y)$ with respect to a parameter vector $\bm \eta_k$ 
in the $k$-th block.
By the definition of mutual information, we have
\begin{equation}
I(X; Y) = h(Y) - h(Y|X),
\end{equation}
where $h(\cdot)$ denotes the differential entropy.
The gradient can therefore be decomposed into two terms:
\begin{equation}
\nabla_{\bm \eta_k} I(X; Y) = 
\nabla_{\bm \eta_k} h(Y) - \nabla_{\bm \eta_k} h(Y|X).
\label{eq:grad_decomp}
\end{equation}

First, let us consider the gradient of the marginal entropy 
$h(Y) = - \mathbb{E}[\log p_{Y}(Y)]$.
It is important to note that the final output $Y$ is a random variable 
whose value depends on the parameter $\bm \eta_k$ through 
the deterministic mappings in the cascade.
We can view $Y$ as a differentiable function of the input $X$, 
all noise variables $\{Z_i\}_{i=1}^N$, 
and the parameters $\{\bm \eta_i\}_{i=1}^N$.
Applying the chain rule to the expectation (assuming standard regularity conditions that allow interchanging differentiation and expectation), we obtain:
\begin{align}
\nabla_{\bm \eta_k} h(Y) 
&= - \mathbb{E}\left[ \nabla_{\bm \eta_k} \log p_{Y}(Y) \right] \nonumber \\
&= - \mathbb{E}\left[ (D_{\bm \eta_k} Y)^\top 
\nabla_{\bm y} \log p_{Y}(\bm y)\big|_{\bm y=Y} \right] \nonumber \\
&\quad - \mathbb{E}\left[ \frac{\partial}{\partial \bm \eta_k} 
\log p_{Y}(\bm y; \bm \eta_k)\big|_{\bm y=Y} \right],
\end{align}
where $D_{\bm \eta_k} Y \equiv \frac{\partial Y}{\partial \bm \eta_k}$ 
is the Jacobian matrix of $Y$ with respect to $\bm \eta_k$.
The second term on the right-hand side vanishes because 
\begin{align}
\mathbb{E}_{Y}\left[\nabla_{\bm \eta_k} \log p_{Y}(Y; \bm \eta_k)\right]
&= \int \nabla_{\bm \eta_k} p_{Y}(\bm y; \bm \eta_k) \,  d\bm y \\
&= \nabla_{\bm \eta_k} \int p_{Y}(\bm y) \, d\bm y \\
&= \nabla_{\bm \eta_k} 1 = 0.
\end{align}
This equation reflects the well-known property that the expectation 
of the score function of the marginal distribution is zero.
By introducing the \textit{marginal score function} 
$s_{Y}(\bm y) \equiv \nabla_{\bm y} \log p_{Y}(\bm y)$, we have
\begin{equation}
\nabla_{\bm \eta_k} h(Y) = - \mathbb{E}\left[ (D_{\bm \eta_k} Y)^\top s_{Y}(Y) \right].
\label{eq:grad_marginal}
\end{equation}

Similarly, for the conditional entropy 
\begin{equation}
h(Y|X) = - \mathbb{E}[\log p_{Y|X}(Y|X)],
\end{equation}
we can derive its gradient by conditioning on $X$.
Let 
\begin{equation}
s_{Y|X}(\bm y | \bm x) \equiv \nabla_{\bm y} \log p_{Y|X}(\bm y | \bm x) 
\end{equation}
be the \textit{conditional score function}.
Following the same logic as above, we obtain:
\begin{equation}
\nabla_{\bm \eta_k} h(Y|X) = - \mathbb{E}\left[ (D_{\bm \eta_k} Y)^\top s_{Y|X}(Y|X) \right].
\label{eq:grad_conditional}
\end{equation}

Substituting \eqref{eq:grad_marginal} and \eqref{eq:grad_conditional} into \eqref{eq:grad_decomp} yields the following key result.

\begin{theorem}[Information Gradient for Cascaded Channels]
\label{thm:cascade_gradient}
The gradient of the end-to-end mutual information $I(X; Y)$ 
with respect to the parameter $\bm \eta_k$ of the $k$-th ($k \in \{1, \dots, N\}$) 
block is given by
\begin{equation}
\nabla_{\bm \eta_k} I(X; Y) = \mathbb{E}\left[ (D_{\bm \eta_k} Y)^\top 
(s_{Y|X}(Y|X) - s_{Y}(Y)) \right].
\label{eq:cascade_info_grad}
\end{equation}
\end{theorem}

This formula has an intuitive interpretation: the gradient is 
the expected product of the Jacobian $D_{\bm \eta_k} Y$, 
which represents how the parameter $\bm \eta_k$ affects the final output, 
and a ``score difference'' 
term $(s_{Y_N|X} - s_{Y_N})$, which represents 
the information-theoretic learning signal.

\begin{remark}[Posterior Score Representation]
  \label{rem:posterior_score}
  \rm 
  A crucial observation is that the score difference term 
  can be equivalently expressed using the posterior distribution:
  \begin{equation}
  s_{Y|X}(\bm y | \bm x) - s_{Y}(\bm y) = \nabla_{\bm y} \log p_{X|Y}(\bm x | \bm y).
  \label{eq:score_diff_posterior}
  \end{equation}
  This identity follows readily from Bayes' rule, 
  $\log p(\bm y|\bm x) - \log p(\bm y) 
  = \log p(\bm x|\bm y) - \log p(\bm x)$, 
  by taking the gradient with respect to $\bm y$ (noting that $\nabla_{\bm y} 
  \log p(\bm x) = 0$).
  Consequently, the information gradient can be re-expressed as
  \begin{equation}
  \nabla_{\bm \eta_k} I(X; Y) = \mathbb{E} 
  \left[ (D_{\bm \eta_k} Y)^\top \nabla_{\bm y} 
  \log p_{X|Y}(X|\bm y)\big|_{\bm y=Y} \right].
  \end{equation}
  This ``posterior score'' form is particularly useful for implementations 
  that utilize an approximate inference model.
\end{remark}

\begin{remark}[Reduction to Single-Stage Gradient]
  \rm It is illuminating to consider the simplest case, 
  a single-stage channel ($N=1$), given by $Y_1 = f_{\bm \eta_1}(X) + Z_1$.
  In this case, the conditional entropy $h(Y_1|X) = h(Z_1)$ is constant 
  with respect to $\bm \eta_1$, implying $\nabla_{\bm \eta_1} h(Y_1|X) = 0$.
  Consequently, the conditional score term in \eqref{eq:cascade_info_grad} 
  vanishes in expectation
  \begin{equation}
  \mathbb{E}[(D_{\bm \eta_1} Y_1)^\top s_{Y_1|X}(Y_1|X)] = 0.
  \end{equation}
  Theorem \ref{thm:cascade_gradient} then reduces to
  \begin{equation}
  \nabla_{\bm \eta_1} I(X; Y_1) 
  = - \mathbb{E}\left[ (D_{\bm \eta_1} Y_1)^\top s_{Y_1}(Y_1) \right],
  \end{equation}
  which perfectly recovers the information gradient formula for 
  standard nonlinear Gaussian channels derived in \cite{wadayama2025-b}.
  Our generalized theorem reveals that for deeper cascades ($N > 1$), 
  the additional term $s_{Y_N|X}$ becomes necessary because intermediate 
  noise injections make $h(Y_N|X)$ dependent on earlier parameters.
\end{remark}

\subsection{Efficient Gradient Computation}

The formula derived in Theorem \ref{thm:cascade_gradient} involves 
the Jacobian matrix $D_{\bm \eta_k} Y$, which can be computationally 
expensive if explicitly instantiated, especially for high-dimensional systems.
However, by exploiting the cascade structure and modern automatic 
differentiation (AD) techniques, we can compute this gradient efficiently 
without ever forming the full Jacobian matrix.

\subsubsection{Chain Rule for Cascaded Jacobian}

Due to the Markovian structure of the cascade, the Jacobian $D_{\bm \eta_k} Y$ 
can be decomposed using the chain rule.
Recall that $Y_N$ depends on $\bm \eta_k$ only through the intermediate 
sequence $Y_k, Y_{k+1}, \dots, Y_{N-1}$.
The Jacobian can be explicitly written as a product 
of Jacobians of purely deterministic mappings:
\begin{equation}
D_{\bm \eta_k} Y = \left( \prod_{i=N}^{k+1} \frac{\partial f_{\bm \eta_i}}{\partial Y_{i-1}}(Y_{i-1}) \right) \cdot \frac{\partial f_{\bm \eta_k}}{\partial \bm \eta_k}(Y_{k-1}).
\label{eq:jacobian_chain_rule}
\end{equation}
Here, $\frac{\partial f_{\bm \eta_i}}{\partial Y_{i-1}}$ is 
the Jacobian of the $i$-th block's function with respect to its input, 
and $\frac{\partial f_{\bm \eta_k}}{\partial \bm \eta_k}$ is 
the Jacobian with respect to its parameter.
This product structure perfectly matches the backward pass of 
standard backpropagation algorithms used in training deep neural networks.

\subsubsection{Vector-Jacobian Product (VJP)}
More importantly, the gradient formula \eqref{eq:cascade_info_grad} does not 
require the full Jacobian matrix itself, but rather its transpose multiplied by a vector.
Let us define the \textit{information score vector} $\bm v \in \mathbb{R}^{d_N}$ as
\begin{equation}
\bm v(\bm x, \bm y) \equiv s_{Y|X}(\bm y | \bm x) - s_{Y}(\bm y).
\end{equation}
The gradient term inside the expectation is then $(D_{\bm \eta_k} Y)^\top \bm v$.
This operation is known as a vector-Jacobian product (VJP) \cite{Baydin2018}.
In standard AD frameworks (like PyTorch, TensorFlow, or JAX), VJP can be computed efficiently 
in a single backward pass by treating $\bm v$ as the initial gradient 
(or ``adjoint'') at the output layer.

In practice, the VJP can be effortlessly computed 
by defining a scalar surrogate loss function.
Let $\text{stop}(\cdot)$ denote the stop-gradient operator, 
which acts as the identity function during the forward pass 
but has zero gradient during the backward pass in the AD framework.
We define the VJP loss as
\begin{equation}
\mathcal{L}_{\text{VJP}}(\bm \eta) \equiv - \langle \text{stop}(\bm v), Y \rangle,
\label{eq:vjp_loss}
\end{equation}
where $\langle \cdot, \cdot \rangle$ denotes the inner product.
Applying standard automatic differentiation to this loss yields
\begin{equation}  
\nabla_{\bm \eta_k} \mathcal{L}_{\text{VJP}} = - (D_{\bm \eta_k} Y)^\top \bm v.
\label{eq:vjp_grad}
\end{equation}
The expectation of this term exactly recovers the negative of 
the information gradient derived in \eqref{eq:cascade_info_grad}.
In practical implementations, we can obtain a Monte Carlo estimator of 
the true information gradient by taking the sample mean 
of $\nabla_{\bm \eta_k} \mathcal{L}_{\text{VJP}}$ over a mini-batch.
For a mini-batch of $B$ samples $\{(\bm x^{(b)}, \bm y^{(b)})\}_{b=1}^B$, 
a Monte Carlo estimator of the information gradient is given by
\begin{equation}
\widehat{\nabla_{\bm \eta_k} I} = - \frac{1}{B} \sum_{b=1}^B 
\nabla_{\bm \eta_k} \mathcal{L}_{\text{VJP}}(\bm \eta_k; \bm x^{(b)}, \bm y^{(b)}).
\label{eq:mc_estimator}
\end{equation}
This estimator allows for maximizing mutual information 
via standard stochastic gradient descent (SGD).

\subsection{Score Model Approximation}

The information gradient formula \eqref{eq:cascade_info_grad} relies on 
the true marginal score $s_Y(\bm y)$ and conditional score 
$s_{Y|X}(\bm y|\bm x)$, which are generally unknown 
for complex nonlinear cascades.
To make the proposed framework practical, we employ parameterized score models, 
typically deep neural networks, to approximate these functions.

Let $s_{\bm \phi}(\bm y; \bm \eta)$ and $s_{\bm \psi}(\bm y|\bm x; \bm \eta)$ 
be the score models parameterized by $\bm \phi$ and $\bm \psi$, 
respectively, designed to approximate the true scores:
\begin{align}
s_{\bm \phi}(\bm y; \bm \eta) &\approx \nabla_{\bm y} \log p_{Y}(\bm y; \bm \eta), \\
s_{\bm \psi}(\bm y|\bm x; \bm \eta) &\approx \nabla_{\bm y} \log p_{Y|X}(\bm y|\bm x; \bm \eta).
\end{align}
Note that these models implicitly depend on the cascade parameters 
$\bm \eta = \{\bm \eta_i\}_{i=1}^N$ since the distributions of $Y$ 
change with $\bm \eta$.
Crucially, these score models can be efficiently trained from samples 
using DSM \cite{vincent2011, song2019, song2021} or 
its variants, without requiring knowledge of the underlying likelihoods.
For standard Gaussian noise channels at the final stage, 
DSM provides a simple and tractable objective.

Replacing the true scores with their learned counterparts 
yields a practical surrogate VJP loss for a sample $(\bm x, \bm y)$:
\begin{equation}
\mathcal{L}_{\text{VJP}}(\bm \eta_k; \bm x, \bm y) 
\equiv - \langle \text{stop}(s_{\bm \psi}(\bm y|\bm x) - s_{\bm \phi}(\bm y)), \bm y \rangle.
\label{eq:surrogate_vjp_loss}
\end{equation}
Optimizing the cascade parameters $\bm \eta$ using gradients derived from this surrogate loss, while simultaneously (or alternately) training the score models $\bm \phi, \bm \psi$ to track the changing distributions, constitutes the core of our proposed end-to-end optimization framework.

\subsection{Analytical Example: Linear Gaussian Cascaded Channel}
To validate the proposed information gradient formula 
(Theorem \ref{thm:cascade_gradient}),
we consider a simple two-stage linear Gaussian cascade where the exact gradient is analytically tractable.
Let the input be $X \sim \mathcal{N}(0, \sigma_X^2)$. Consider a two-stage cascade ($N=2$) with scalar variables:
\begin{align}
Y_1 &= \eta_1 X + Z_1, \quad Z_1 \sim \mathcal{N}(0, \sigma_1^2), \\
Y_2 &= Y_1 + Z_2, \quad \quad Z_2 \sim \mathcal{N}(0, \sigma_2^2),
\end{align}
where $\eta_1$ is the scalar parameter of interest in the first stage.
The total effective noise is $Z_{eff} = Z_1 + Z_2 \sim \mathcal{N}(0, \sigma_N^2)$ 
with $\sigma_N^2 = \sigma_1^2 + \sigma_2^2$.
The final output is $Y_2 = \eta_1 X + Z_{eff}$, 
with variance $v_{Y_2} \equiv \text{Var}(Y_2) = \eta_1^2 \sigma_X^2 + \sigma_N^2$.

The mutual information for this scalar Gaussian channel is known to be
\begin{equation}
I(X; Y_2) = \frac{1}{2} \log \left( 1 + \frac{\eta_1^2 \sigma_X^2}{\sigma_N^2} \right) = \frac{1}{2} \log \left( \frac{v_{Y_2}}{\sigma_N^2} \right).
\end{equation}
Differentiating this directly with respect to $\eta_1$ yields the true gradient:
\begin{equation}
\frac{\partial I(X; Y_2)}{\partial \eta_1} = \frac{1}{2} \frac{\sigma_N^2}{v_{Y_2}} \cdot \frac{2 \eta_1 \sigma_X^2}{\sigma_N^2} = \frac{\eta_1 \sigma_X^2}{v_{Y_2}}.
\label{eq:true_grad_linear}
\end{equation}

%\subsection{Gradient via Proposed Formula}
Now we apply Theorem \ref{thm:cascade_gradient}.
The Jacobian of the final output $Y_2$ with respect to $\eta_1$ is 
$D_{\eta_1} Y_2 = \frac{\partial}{\partial \eta_1}(\eta_1 X + Z_1 + Z_2) = X$.
The marginal score is $s_{Y_2}(y) = - y / v_{Y_2}$, and the conditional score is $s_{Y_2|X}(y|x) = - (y - \eta_1 x) / \sigma_N^2$.
Substituting these into \eqref{eq:cascade_info_grad}:
\begin{align}
\nabla_{\eta_1} I &= \mathbb{E}\left[ X \cdot \left( -\frac{Y_2 - \eta_1 X}{\sigma_N^2} - \left(-\frac{Y_2}{v_{Y_2}}\right) \right) \right] \nonumber \\
&= -\frac{\mathbb{E}[X Y_2] - \eta_1 \mathbb{E}[X^2]}{\sigma_N^2} + \frac{\mathbb{E}[X Y_2]}{v_{Y_2}}.
\end{align}
Using $\mathbb{E}[X^2] = \sigma_X^2$ and $\mathbb{E}[X Y_2] = \mathbb{E}[X(\eta_1 X + Z_{eff})] = \eta_1 \sigma_X^2$, the first term vanishes:
\begin{equation}
-\frac{\eta_1 \sigma_X^2 - \eta_1 \sigma_X^2}{\sigma_N^2} + \frac{\eta_1 \sigma_X^2}{v_{Y_2}} = \frac{\eta_1 \sigma_X^2}{v_{Y_2}}.
\end{equation}
This exactly matches the classical result in \eqref{eq:true_grad_linear}, confirming the validity of our generalized information gradient formula.

\section{Generalization to Directed Acyclic Graphs}

Having established the information gradient for cascaded channels, 
we now extend our framework to general stochastic networks represented 
by directed acyclic graphs (DAGs). This generalization allows us to optimize 
complex systems involving branching, merging, and parallel processing paths.

\subsection{Problem Setup}

Consider a communication or sensing system represented by 
a directed acyclic graph (DAG) $\mathcal{G} = (\mathcal{V}, \mathcal{E})$ with $M$ nodes.
Let $V_1, \dots, V_M$ denote the random vectors associated with the nodes in a topological order.
The network input is represented by a subset of root nodes, collectively denoted by $X$.
Each subsequent node $V_j$ is computed from its parent nodes $\text{Pa}(V_j)$ via 
a parameterized deterministic mapping, optionally corrupted by independent additive noise.
Specifically, the local computation at node $V_j$ is given by
\begin{equation}
V_j = f_j(\text{Pa}(V_j); \bm \eta_j) + Z_j,
\label{eq:local_computation}
\end{equation}
where $f_j:\mathbb{R}^{d_{j}^{\text{in}}} \to \mathbb{R}^{d_{j}^{\text{out}}}$ is a differentiable function parameterized by $\bm \eta_j$, 
and $Z_j$ is a zero-mean independent Gaussian noise vector (if node $V_j$ is deterministic, $Z_j \equiv 0$).
The set of all network parameters is $\bm \eta \equiv \{\bm \eta_j\}_j$, 
and the set of all noise sources is $\mathcal{Z} \equiv \{Z_j\}_j$.

\begin{figure}[htbp]
  \centering
  \includegraphics[width=\linewidth]{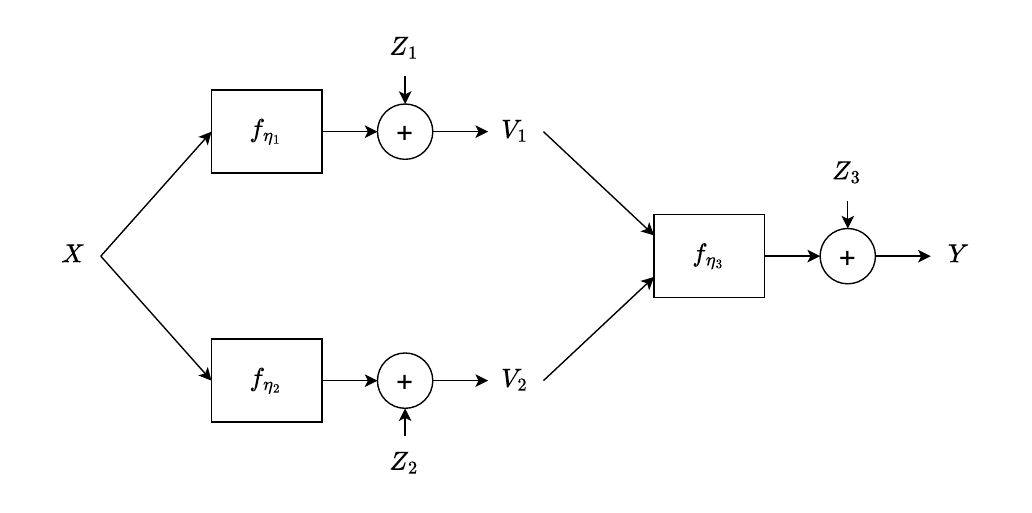}
  \caption{An illustrative DAG network. The input $X$ branches into two parallel paths, 
  each processed by a distinct function $f_{\bm \eta_i}$ and corrupted by noise $Z_i$ to 
  produce intermediate nodes $V_1, V_2$. These are then merged by function $f_{\bm \eta_3}$ 
  and further corrupted by noise $Z_3$ to yield the final output $Y$.}
  \label{fig:dag_network}
\end{figure}

The final output of the system, denoted by $Y$, is a subset of the leaf nodes.
Thanks to the acyclic structure of $\mathcal{G}$, recursively substituting the parent nodes 
in \eqref{eq:local_computation} is guaranteed to terminate.
Consequently, the end-to-end relationship from the roots $X$ to the outputs $Y$ can be 
explicitly expressed as a deterministic composition of the input $X$ and all involved noise sources $\mathcal{Z}$:
\begin{equation}
Y = g(X, \mathcal{Z}; \bm \eta).
\label{eq:dag_mapping}
\end{equation}
Figure \ref{fig:dag_network} shows a schematic representation of a DAG network.
To ensure the existence of the necessary score functions, 
we assume sufficient noise is injected into the network (typically at the final output nodes) 
such that $Y$ admits a differentiable probability density function.

\subsection{Information Gradient Formula for DAG channels}

We now derive the gradient of the mutual information $I(X; Y)$ with respect to an arbitrary parameter vector $\bm \eta_k \in \bm \eta$ within the DAG.
Remarkably, the gradient formula retains the same elegant structure as in the cascade case, demonstrating its universality.

\begin{theorem}[Information Gradient for DAGs]
\label{thm:dag_gradient}
Under standard regularity conditions, the gradient of the end-to-end mutual information $I(X; Y)$ with respect to any internal parameter $\bm \eta_k$ in a DAG is given by
\begin{equation}
\nabla_{\bm \eta_k} I(X; Y) = \mathbb{E}\left[ (D_{\bm \eta_k} Y)^\top (s_{Y|X}(Y|X) - s_{Y}(Y)) \right],
\label{eq:dag_info_grad}
\end{equation}
where $D_{\bm \eta_k} Y \equiv \frac{\partial g(X, \mathcal{Z}; \bm \eta)}{\partial \bm \eta_k}$ is the 
Jacobian of the final output with respect to $\bm \eta_k$, 
and the expectation is taken over all random 
variables $(X, \mathcal{Z})$ in the network.
\end{theorem}

\begin{IEEEproof}
The mutual information is $I(X; Y) = h(Y) - h(Y|X)$.
We analyze the gradient of the marginal entropy $h(Y) = - \mathbb{E}[\log p_Y(Y)]$ first.
Unlike the cascade case, the expectation is now taken over 
the joint distribution of all inputs and noise sources in the DAG.
Let $\bm \omega = (X, \mathcal{Z})$ denote the collection of all primitive random variables in the network.
The output $Y$ is a deterministic function $Y = g(\bm \omega; \bm \eta)$.
Assuming regularity conditions that permit interchanging $\nabla_{\bm \eta_k}$ and $\mathbb{E}_{\bm \omega}$, we have:
\begin{align}
\nabla_{\bm \eta_k} h(Y) &= - \mathbb{E}_{\bm \omega} \left[ \nabla_{\bm \eta_k} \log p_Y(g(\bm \omega; \bm \eta); \bm \eta) \right] \\
&= - \mathbb{E}_{\bm \omega} \left[ (D_{\bm \eta_k} Y)^\top \nabla_{\bm y} \log p_Y(\bm y)\big|_{\bm y=Y} \right] \nonumber \\
&\quad - \mathbb{E}_{\bm \omega} \left[ \frac{\partial}{\partial \bm \eta_k} \log p_Y(\bm y; \bm \eta)\big|_{\bm y=Y} \right].
\end{align}
The second term is $\mathbb{E}_Y[\nabla_{\bm \eta_k} \log p_Y(Y; \bm \eta)]$, which vanishes due to the score function property $\int \nabla p = \nabla \int p = 0$.
Thus, $\nabla_{\bm \eta_k} h(Y) = - \mathbb{E}\left[ (D_{\bm \eta_k} Y)^\top s_Y(Y) \right]$.
Following a strictly analogous argument for the conditional entropy term (conditioning on $X$ throughout), we obtain $\nabla_{\bm \eta_k} h(Y|X) = - \mathbb{E}\left[ (D_{\bm \eta_k} Y)^\top s_{Y|X}(Y|X) \right]$.
Combining these results yields the claim of the theorem.
\end{IEEEproof}

Crucially, the Jacobian $D_{\bm \eta_k} Y$ in a DAG can be efficiently computed using standard reverse-mode automatic differentiation (backpropagation) through the graph, exactly as in modern deep learning frameworks.
This means the VJP-based efficient computation strategy described in Section III-C applies directly to general DAGs without modification.

\begin{remark}[Mathematical Importance of Acyclicity]
  \rm The assumption that the graph $\mathcal{G}$ is acyclic (DAG) is fundamental to our framework for two key reasons:
  \begin{enumerate}
      \item \textbf{Well-defined Composition:} The absence of cycles guarantees that the recursive computation terminates in finite steps. This ensures that the end-to-end mapping $Y = g(X, \mathcal{Z}; \bm \eta)$ is a strictly well-defined composite function. If cycles were present, the system would be self-referential, requiring fixed-point formulations.
      \item \textbf{Applicability of Chain Rule:} Because $Y$ is a standard composite function, its Jacobian $D_{\bm \eta_k} Y$ is rigorously guaranteed to exist (under local differentiability assumptions) and can be computed via the standard multivariate chain rule. This theoretical guarantee legitimizes the use of backpropagation (VJP) for efficient computation.
  \end{enumerate}
\end{remark}

\begin{remark}[Reparameterization and Differentiability]
  \rm 
  \label{rem:reparam}
    A natural question arises regarding the differentiability 
    of the stochastic output 
    $Y$ with respect to parameters $\bm \eta$. Our framework relies on 
    the \textit{reparameterization trick} \cite{Kingma2014} (also known as the pathwise derivative estimator). 
    As explicitly formulated in \eqref{eq:dag_mapping} 
    , we treat the system output not as a mere random variable, but as a deterministic function 
    $Y=g(X, \mathcal{Z}; \eta)$ of the parameters and fixed independent noise sources $\mathcal{Z}$. 
    Since the randomness is externalized into $\mathcal{Z}$ which does not depend on $\bm \eta$, 
    the Jacobian $D_{\eta}Y$ is well-defined for any fixed realization of $(X, \mathcal{Z})$. 
    This perspective effectively decouples the randomness from the parametric dependence, 
    allowing the gradient to ``flow'' through the stochastic nodes via standard backpropagation.
  \end{remark}

\begin{remark}[Generality of Noise Assumptions]
  \rm
  While this paper primarily assumes Gaussian noise for efficient score matching (DSM), 
  the derived information gradient formulas theoretically hold for {any noise distribution} that 
  admits a differentiable probability density function. Furthermore, {noise injection is not required 
  at every node}. As long as sufficient noise enters the DAG (typically at the final output or 
  accumulated along the path) to ensure the output $Y$ admits a smooth density, 
  intermediate transformations can be purely deterministic ($Z_j \equiv 0$). 
  This flexibility allows our framework to model systems with mixed stochastic 
  and deterministic components, such as digital logic combined with analog transmission.
\end{remark}

\subsection{Analytical Example: Multipath DAG}
\label{subsec:multipath-dag}
To further validate the generalized information gradient for DAGs 
(Theorem \ref{thm:dag_gradient}), we consider a multipath network where parameters affect the output through different topological structures.
Consider the DAG shown in Fig. \ref{fig:dag_network} with the following linear Gaussian specifications:
\begin{align}
    V_1 &= \eta_1 X + Z_1, \quad Z_1 \sim \mathcal{N}(0, \sigma_1^2) \\
    V_2 &= X + Z_2, \quad \quad Z_2 \sim \mathcal{N}(0, \sigma_2^2) \\
    Y &= \eta_2 V_1 + V_2 + Z_3, \quad Z_3 \sim \mathcal{N}(0, \sigma_3^2)
\end{align}
where $X \sim \mathcal{N}(0, \sigma_X^2)$.
Here, $\eta_1$ is a deep parameter affecting only one path, while $\eta_2$ is a mixing parameter at the merge node.
Crucially, $\eta_2$ affects both the signal gain and the effective noise variance.
The end-to-end relationship is $Y = G_{\text{eff}} X + Z_{\text{eff}}$, where $G_{\text{eff}} = \eta_1\eta_2 + 1$ is the effective gain and $Z_{\text{eff}} = \eta_2 Z_1 + Z_2 + Z_3$ is the effective noise with variance $\sigma_N^2 = \eta_2^2\sigma_1^2 + \sigma_2^2 + \sigma_3^2$.
The output variance is $v_Y = G_{\text{eff}}^2 \sigma_X^2 + \sigma_N^2$.

The mutual information is  simply given by  $I(X;Y) = \frac{1}{2}\log(v_Y/\sigma_N^2)$.
Direct differentiation yields the true analytical gradients:
\begin{align}
    \frac{\partial I}{\partial \eta_1} &= \frac{\eta_2 G_{\text{eff}} \sigma_X^2}{v_Y} \label{eq:true_grad_eta1} \\
    \frac{\partial I}{\partial \eta_2} &= \frac{\eta_1 G_{\text{eff}} \sigma_X^2 + \eta_2 \sigma_1^2}{v_Y} - \frac{\eta_2 \sigma_1^2}{\sigma_N^2} \label{eq:true_grad_eta2}
\end{align}
We now apply Theorem \ref{thm:dag_gradient}. The Jacobians $D_{\eta_k}Y$ are derived via the chain rule over the DAG:
\begin{equation}
    D_{\eta_1}Y = \eta_2 X, \quad D_{\eta_2}Y = V_1 = \eta_1 X + Z_1.
\end{equation}
Using the Gaussian scores $s_Y(y) = -y/v_Y$ and $s_{Y|X}(y|x) = -(y - G_{\text{eff}}x)/\sigma_N^2 = -Z_{\text{eff}}/\sigma_N^2$, we verify the theorem:
For $\eta_1$, since $\mathbb{E}[X Z_{\text{eff}}] = 0$:
\begin{equation}
    \nabla_{\eta_1}I = \mathbb{E}\left[\eta_2 X \left( s_{Y|X} - s_Y \right)\right] = \eta_2 \mathbb{E}\left[ X \frac{Y}{v_Y} \right] = \frac{\eta_2 G_{\text{eff}}\sigma_X^2}{v_Y}.
\end{equation}
This exactly matches \eqref{eq:true_grad_eta1}.
For $\eta_2$, the Jacobian $V_1$ is correlated with $Z_{\text{eff}}$ via $Z_1$:
\begin{align}
    \nabla_{\eta_2}I &= \mathbb{E}\left[ V_1 \left( -\frac{Z_{\text{eff}}}{\sigma_N^2} + \frac{Y}{v_Y} \right) \right] \nonumber \\
    &= -\frac{\mathbb{E}[(\eta_1 X + Z_1)(\eta_2 Z_1 + \dots)]}{\sigma_N^2} + \frac{\mathbb{E}[V_1 Y]}{v_Y} \nonumber \\
    &= -\frac{\eta_2 \sigma_1^2}{\sigma_N^2} + \frac{\eta_1 G_{\text{eff}} \sigma_X^2 + \eta_2 \sigma_1^2}{v_Y}.
\end{align}
This perfectly recovers \eqref{eq:true_grad_eta2}.
This confirms that our framework correctly handles complex dependencies where parameters simultaneously affect signal paths and noise characteristics.

\subsection{Algorithm for Information Gradient Estimation}

The information gradient formula \eqref{eq:dag_info_grad} allows us to estimate 
the information gradient for any parameter in the DAG using a Monte Carlo approach.
Algorithm \ref{alg:gradient_estimation} outlines the procedure 
to obtain a mini-batch estimate of the gradient $\nabla_{\bm \eta_k} I(X; Y)$ 
for a target parameter $\bm \eta_k$.
We assume that the necessary score models $s_{\bm \phi}(\bm y) \approx 
\nabla_{\bm y} \log p_Y(\bm y)$ and $s_{\bm \psi}(\bm y|\bm x) \approx \nabla_{\bm y} \log p_{Y|X}(\bm y|\bm x)$ 
have been pre-trained or are being trained concurrently.

\begin{algorithm}[h]
\caption{Information Gradient Estimation via VJP}
\label{alg:gradient_estimation}
\begin{algorithmic}[1]
\Require DAG $\mathcal{G}$, target parameter $\bm \eta_k$, learned score models $s_{\bm \phi}, s_{\bm \psi}$, mini-batch of inputs $\{\bm x^{(i)}\}_{i=1}^B$.
\Ensure Gradient estimate $\widehat{\nabla_{\bm \eta_k} I}$.

\State \textbf{// 1. Forward Pass}
\For{$i = 1$ to $B$}
    \State Run foward sampling on the DAG from input $\bm x^{(i)}$ to obtain output $\bm y^{(i)}$ and all intermediate node values.
\EndFor

\State \textbf{// 2. Score Difference Calculation}
\For{$i = 1$ to $B$}
    \State Compute marginal score: $\bm s^{(i)}_{\text{marg}} \leftarrow s_{\bm \phi}(\bm y^{(i)})$
    \State Compute conditional score: $\bm s^{(i)}_{\text{cond}} \leftarrow s_{\bm \psi}(\bm y^{(i)} | \bm x^{(i)})$
    \State Compute score difference vector: $\bm v^{(i)} \leftarrow \bm s^{(i)}_{\text{cond}} - \bm s^{(i)}_{\text{marg}}$
\EndFor

\State \textbf{// 3. Backward Pass (VJP)}
\State Define scalar VJP loss: $\mathcal{L} \leftarrow - \frac{1}{B} \sum_{i=1}^B \langle \text{stop}(\bm v^{(i)}), \bm y^{(i)} \rangle$
\State Run backpropagation from $\mathcal{L}$ through the DAG.
\State Extract gradient for target parameter: $\widehat{\nabla_{\bm \eta_k} I} \leftarrow - \nabla_{\bm \eta_k} \mathcal{L}$
\end{algorithmic}
\end{algorithm}

This estimation procedure efficiently computes the gradients 
for {\em all} parameters in the DAG simultaneously in a single backward pass, 
not just for one specific $\bm \eta_k$.

\subsection{Integral Representation of Mutual Information}

\subsubsection{Path Integral Representation}
While our primary focus is gradient-based optimization, 
the proposed framework also enables the estimation of the mutual information value $I(X; Y)$ itself.
By leveraging the fundamental theorem of line integrals, we can reconstruct the mutual information 
from its gradients:
\begin{equation}
I(X; Y; \bm \eta_1) = I(X; Y; \bm \eta_0) + \int_{0}^{1} \langle \nabla_{\bm \eta} I(\bm \eta(t)), \dot{\bm \eta}(t) \rangle dt,
\label{eq:path_integral}
\end{equation}
where $\bm \eta(t)$ is any smooth path in the parameter space connecting 
a reference configuration $\bm \eta_0 = \bm \eta(0)$ to the target configuration $\bm \eta_1 = \bm \eta(1)$.
In many practical scenarios, we can choose a reference $\bm \eta_0$ where the mutual information 
is known to be zero (e.g., by setting parameters to zero such that the signal is completely blocked).
By numerically integrating the gradient estimates obtained via Algorithm \ref{alg:gradient_estimation} 
along such a path (e.g., using the trapezoidal rule), we can estimate the absolute value of 
the end-to-end mutual information for any general DAG.

\subsubsection{Fisher Integral Representation}
The mutual information $I(X; Y)$ can be directly estimated 
by utilizing the connection between differential entropy and 
Fisher information, known as de Bruijn's identity.
Let $Y_0 \equiv Y$ be the DAG output, and consider a noise-corrupted 
version $Y_t = Y_0 + \sqrt{t} Z'$, where $Z' \sim \mathcal{N}(0, I)$ is 
an independent standard Gaussian noise vector and $t \ge 0$ is the added noise variance.
Recall that the mutual information is the difference of differential entropies:
$I(X; Y_t) = h(Y_t) - h(Y_t|X)$.
De Bruijn's identity states that the derivative of differential entropy with respect to the noise variance $t$ is proportional to the Fisher information:
\begin{equation}
\frac{d}{dt} h(Y_t) = \frac{1}{2} J(Y_t), \quad \frac{d}{dt} h(Y_t|X) = \frac{1}{2} J(Y_t|X),
\end{equation}
where $J(Y_t) \equiv \mathbb{E}[\|\nabla_{\bm y_t} \log p(Y_t) \|^2]$ 
and $J(Y_t|X) \equiv \mathbb{E}[\| \nabla_{\bm y_t} \log p(Y_t|X) \|^2]$ 
are the unconditional and conditional Fisher information, respectively.
Consequently, the derivative of mutual information is given by:
\begin{equation}
\frac{d}{dt} I(X; Y_t) = \frac{1}{2} \left( J(Y_t) - J(Y_t|X) \right).
\label{eq:de_bruijn}
\end{equation}
Using the property that the conditional expectation of the conditional score 
is the marginal score, i.e., $\mathbb{E}[s_{Y_t|X}(Y_t|X) | Y_t] = s_{Y_t}(Y_t)$, 
we have the identity
\begin{equation}
J(Y_t|X) - J(Y_t) = \mathbb{E}\left[ \| s_{Y_t|X}(Y_t|X) - s_{Y_t}(Y_t) \|^2 \right].
\end{equation}
Substituting this into \eqref{eq:de_bruijn} and 
integrating from $t=0$ to $\infty$ (noting that $\lim_{t\to\infty} I(X; Y_t) = 0$), 
we obtain the following integral representation:
\begin{align}
I(X; Y) 
&= \frac{1}{2} \int_0^\infty [J(Y_t|X) - J(Y_t)] dt \\
&= \frac{1}{2} \int_0^\infty \mathbb{E}\left[ \| s_{Y_t|X}(Y_t|X) - s_{Y_t}(Y_t) \|^2 \right] dt.
\label{eq:fisher_integral}
\end{align}
This formula allows for estimating $I(X; Y)$ by training score models 
for varying noise levels $t$ and performing numerical integration.
Notably, this approach does not require a reference parameter configuration $\bm \eta_0$ with 
known mutual information, as required in the path-integral method.
This integral representation is fundamentally grounded 
in the score-to-Fisher bridge (SFB) framework introduced 
in \cite{wadayama2025}.

\subsection{Relation to prior work}
A large body of conventional MI estimation and representation 
learning relies on variational or contrastive bounds 
(e.g., InfoNCE, CLUB, MINE), which optimize a discriminator 
or density-ratio model and thus inherit bias-variance 
and tuning issues from critic training.
Non-parametric methods based on kernel density estimation (KDE) \cite{moon1995}
or $k$-nearest neighbors ($k$NN) \cite{Kraskov2004} are also widely used for 
MI estimation. However, they often suffer from the curse 
of dimensionality and are not naturally suited 
for gradient-based optimization, 
as differentiating through these estimators 
with respect to system parameters is computationally prohibitive.

In contrast, our approach does not 
use a variational lower bound; it expresses the \emph{exact} 
information gradient as a product of a score difference and 
a VJP, making it directly 
amenable to automatic differentiation 
and avoiding adversarial/contrastive objectives. 
Classical identities (I-MMSE \cite{guo2005}, 
de Bruijn's identity \cite{stam1959}) 
relate derivatives of $I(X;Y)$ to MMSE or Fisher information 
in additive Gaussian settings; 
we build on these insights but extend them to general nonlinear 
DAGs and arbitrary internal parameters via a unified gradient 
formula computable with VJPs. When closed-form scores are unavailable, 
we leverage DSM
(and conditional/posterior variants) 
to learn the required scores, in contrast to likelihood-ratio 
or critic-based estimators. 
For estimating MI values (not only gradients), 
our path-integral and Fisher-integral representations 
provide an alternative to density-ratio and variational estimators, 
enabling integration of readily estimated gradients/Fisher terms. 
% Replace placeholders with your bibliography keys if desired:
% e.g., \cite{oord2018infonce,poole2019variational,nguyen2010ratio,belghazi2018mine,hyvarinen2005score,vincent2011dsm,guo2005immse}.

\section{Mutual Information Maximization under Global Constraints}

The derived information gradient enables the optimization of DAG networks under various practical constraints.
In many engineering scenarios, such as power-limited sensor networks or bandwidth-constrained communication systems, 
the network must be optimized under a global resource constraint rather than individual local constraints.

\subsection{Problem Formulation}
We consider the problem of maximizing the end-to-end mutual information subject to a global constraint 
on the network parameters $\bm \eta$.
Let $C(\bm \eta): \mathbb{R}^{\dim(\bm \eta)} \to \mathbb{R}$ be a differentiable convex cost function 
representing the total resource usage of the network (e.g., total power consumption defined by an $L_2$ norm $C(\bm \eta) = \|\bm \eta\|_2^2$).
The optimization problem is formulated as
\begin{align}
\text{maximize}_{\bm \eta} \  I(X; Y) \ \text{subject to} \  C(\bm \eta) \le P, 
\label{eq:opt_prob}
\end{align}
where $P > 0$ is the maximum allowable resource budget.

\subsection{Projected Gradient Ascent}
To solve \eqref{eq:opt_prob}, we can employ the {\em projected gradient ascent} (PGA) method.
The PGA algorithmis a standard iterative optimization algorithm suitable for constrained problems 
where the projection onto the feasible set $\mathcal{S} = \{\bm \eta \mid C(\bm \eta) \le P\}$ 
is computationally tractable.

The update rule at iteration $t$ consists of two steps: a gradient ascent step 
using the estimated information gradient, followed by a projection step onto $\mathcal{S}$:
\begin{align}
\tilde{\bm \eta}^{(t+1)} &\leftarrow \bm \eta^{(t)} + \alpha_t \widehat{\nabla_{\bm \eta} I}(\bm \eta^{(t)}) \\
\bm \eta^{(t+1)} &\leftarrow \mathcal{P}_{\mathcal{S}}(\tilde{\bm \eta}^{(t+1)}).
\end{align}
Here, $\alpha_t > 0$ is the learning rate (step size), and $\widehat{\nabla_{\bm \eta} I}$ 
is the gradient estimated using Algorithm \ref{alg:gradient_estimation} on a mini-batch of data.
The operator $\mathcal{P}_{\mathcal{S}}(\cdot)$ denotes the Euclidean projection onto the feasible set $\mathcal{S}$:
\begin{equation}
\mathcal{P}_{\mathcal{S}}(\bm v) \triangleq \argmin_{\bm \eta' \in \mathcal{S}} \| \bm \eta' - \bm v \|_2.
\end{equation}
For common constraints like the global power constraint ($C(\bm \eta) = \|\bm \eta\|_2^2 \le P$), 
this projection has a simple closed-form solution (scaling the vector if its norm exceeds $\sqrt{P}$).
By repeating these steps, the network parameters $\bm \eta$ are updated 
to maximize the end-to-end information flow while strictly satisfying the global resource constraint.

\begin{remark}[Convergence and Non-convexity]
  \rm In general, the mutual information objective $I(X; Y)$ is 
  non-convex  with respect to the network parameters $\bm \eta$, particularly when 
  the DAG involves complex nonlinear mappings such as deep neural networks.
  Consequently, standard gradient-based optimization methods, 
  including the proposed PGA algorithm, are guaranteed to converge 
  only to a stationary point (local optimum or saddle point) 
  of the objective function, not necessarily to the global maximum.
  Despite this theoretical limitation, gradient-based maximization 
  of MI has been shown to be highly effective 
  in various practical applications.
\end{remark}

\begin{remark}[Mitigation Strategies for Non-Convexity]
  \rm To mitigate the risk of getting trapped in poor local optima, 
  several standard practical strategies can be employed.
  One effective approach is \textit{multi-round optimization}, 
  where the gradient ascent is executed multiple times with different 
  random initializations of $\bm \eta$, and the solution with the highest 
  estimated mutual information is selected.
  Another promising approach is \textit{noise annealing}. 
  By artificially injecting extra Gaussian noise into $Y$ and 
  gradually decreasing its variance to zero during optimization, 
  the objective landscape is initially smoothed, potentially allowing 
  the optimizer to escape shallow local optima before converging 
  to a final solution.
\end{remark}

\subsection{Practical Consideration: Score Drift}
\label{sec:score_drift}
A critical challenge in the implementation of PGA algorithm presented above is that 
updating the network parameters $\bm \eta$ inherently changes the underlying data distributions 
$p_Y(\bm y; \bm \eta)$ and $p_{Y|X}(\bm y|\bm x; \bm \eta)$.
Consequently, the true score functions ``drift'' during the optimization process.
Using a stale score model trained on an old parameter configuration to estimate gradients 
for a new configuration can lead to biased or incorrect update directions.

To mitigate this issue, two main strategies can be employed:
\begin{itemize}
    \item \textbf{Iterative Retraining:} The simplest approach is 
    to alternate between score learning and parameter optimization. 
    After every few gradient ascent steps for $\bm \eta$, the score models are fine-tuned 
    using fresh samples generated from the current DAG.
    \item \textbf{Conditional Score Networks:} A more advanced approach is to 
    train \textit{conditional} score models $s_{\bm \phi}(\bm y; \bm \eta)$ 
    and $s_{\bm \psi}(\bm y|\bm x; \bm \eta)$ that explicitly take the network parameters $\bm \eta$ 
    as additional inputs. By pre-training these models over the relevant range of $\bm \eta$, 
    we can perform optimization without frequent retraining, significantly accelerating the process.
\end{itemize}
Both strategies present a trade-off between implementation complexity and computational efficiency.
Iterative retraining is straightforward to implement and ensures highly accurate 
local score estimates, but it can become a computational bottleneck during the optimization 
phase due to frequent re-training.
In contrast, conditional score networks significantly accelerate the online optimization phase 
by eliminating the need for retraining.
However, they require a more complex and potentially data-intensive pre-training phase to ensure 
the model generalizes well across the relevant parameter space of $\bm \eta$.

\subsection{Potential Impact and Applications}

The proposed framework, which enables end-to-end information-theoretic optimization of general DAGs under global constraints, has significant potential impacts across various domains involving complex stochastic systems.

\subsubsection{Next-Generation Wireless Systems (Semantic Communications)}
Unlike traditional communications that aim to maximize mere data transmission rates, 
next-generation paradigms like \textit{semantic communication} \cite{Gunduz2023}
focus on maximizing 
the successful execution of downstream tasks.
Our framework naturally supports this goal by treating the entire communication link as a DAG.
Crucially, this allows for the 
joint optimization of transmitter (encoder) and receiver (decoder).
By modeling the encoder $\bm \eta_{\text{TX}}$, the noisy channel, 
and the decoder $\bm \eta_{\text{RX}}$ as a single unified DAG, 
we can simultaneously optimize both $\bm \eta_{\text{TX}}$ and $\bm \eta_{\text{RX}}$ 
to maximize the task-relevant mutual information $I(T; Y_{\text{final}})$ at the receiver's decision point, 
rather than just maximizing channel capacity.
This end-to-end approach is particularly valuable when the channel involves complex, 
nonlinear distortions that are difficult to handle with classical modular designs.

\subsubsection{IoT and Distributed Sensor Networks}
Consider a large-scale sensor network where numerous battery-constrained sensors observe 
a common physical phenomenon $X$ and transmit processed data to a central hub $Y$.
This system forms a massive DAG with a global power constraint $C(\{\bm \eta_i\}) \le P_{\text{total}}$.
Our framework provides a principled way to determine the optimal operational policy
(which sensors should be active, how much they should compress data, and how much power they should allocate 
for transmission) to maximize the aggregated information $I(X; Y)$ at the center.

\subsubsection{Deep Learning and Representation Learning}
Deep neural networks (DNNs) can be viewed as deterministic DAGs 
(or stochastic ones if techniques like dropout or noisy activations are used).
Our method offers a direct way to integrate information-theoretic principles, 
such as the \textit{Information Bottleneck (IB)} principle \cite{Tishby1999}, 
into DNN training.
By utilizing the derived gradients for the IB objective $\mathcal{L}_{\text{IB}} = I(T; Y) - \beta I(X; Y)$, 
we can train networks to learn representations $Y$ that are maximally informative 
about a target task $T$ while being minimally informative about the raw input $X$, 
thereby improving generalization and robustness.

\section{Relation to Network Information Theory}

The proposed DAG framework is not merely a tool for arbitrary computational graphs 
but naturally encompasses many fundamental channel models studied in  network information theory \cite{cover2006}.
By viewing these classical models through the lens of DAGs, 
our gradient-based approach provides a unified numerical solver for their end-to-end mutual 
infomation, capacity or achievable rate 
maximization problems, even under complex nonlinear or non-Gaussian assumptions where 
analytical solutions are intractable.

\subsection{Multi-User Channel Models as DAGs}

\begin{figure}[htbp]
  \centering
  \includegraphics[width=0.9\linewidth]{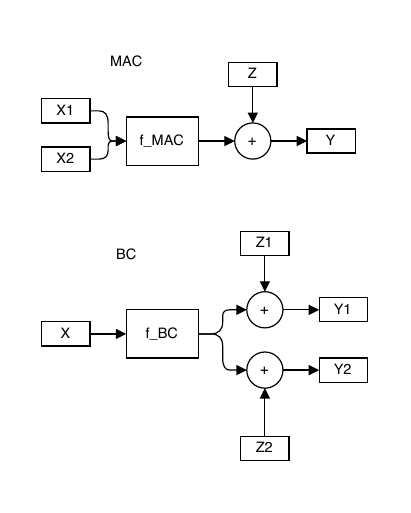}
  \caption{MAC channel model and BC channel model.}
  \label{fig:MAC_channel}
\end{figure}

Many canonical multi-user channels can be directly mapped to specific DAG topologies:

\begin{itemize}
    \item \textbf{Multiple Access Channel (MAC):} A MAC with two senders $X_1, X_2$ and 
    one receiver $Y$ is represented by a DAG where two root nodes merge into a single child node.
    \begin{equation}
    Y = f_{\text{MAC}}(X_1, X_2; \bm \eta) + Z.
    \end{equation}
    Maximizing $I(X_1, X_2; Y)$ under individual power constraints on $X_1, X_2$ yields the sum-capacity point.

    \item \textbf{Broadcast Channel (BC):} A BC with one sender $X$ 
    and two receivers $Y_1, Y_2$ corresponds to a DAG where a single root node branches into two leaf nodes.
    \begin{equation}
    Y_1 = f_1(X; \bm \eta_1) + Z_1, \quad Y_2 = f_2(X; \bm \eta_2) + Z_2.
    \end{equation}
    The capacity region involves trade-offs between $I(X; Y_1)$ and $I(X; Y_2)$, 
    often managed by optimizing the input distribution $p_X$ or superposition coding parameters $\bm \eta$.
    % \item \textbf{Relay Channel:} A classic three-node relay channel 
    % (Source $X$, Relay $X_R/Y_R$, Destination $Y$) forms a DAG with both branching 
    % (broadcast from $X$ to $Y_R, Y$) and merging (multiple access at $Y$ from $X, X_R$).
    % Assuming a parameterized relay function $X_R = f_R(Y_R; \bm \eta_R)$, the system is described by:
    % \begin{align}
    % Y_R &= f_1(X) + Z_1, \\
    % X_R &= f_R(Y_R; \bm \eta_R), \\
    % Y &= f_2(X, X_R) + Z_2.
    % \end{align}
    %Our framework can directly optimize the relay function parameters $\bm \eta_R$ 
    %to maximize the end-to-end mutual information $I(X; Y)$.
\end{itemize}
These two channel models are illustrated in Figure \ref{fig:MAC_channel}.

\subsection{A Unified Numerical Solver}
Traditionally, finding the capacity regions of these models often requires 
model-specific analytical techniques (e.g., entropy inequalities, converse proofs), 
which become extremely difficult when components are nonlinear or non-Gaussian.
Our DAG-based information gradient framework offers a \textit{unified numerical approach}: 
by specifying the graph topology and identifying tunable parameters $\bm \eta$ 
(which can parametrize encoders, decoders, or relay functions), 
one can apply the same generic optimization algorithm (Algorithm \ref{alg:gradient_estimation} + PGA) 
to numerically discover achievable rate regions for a wide variety of network scenarios.

\subsection{Optimization of Input Distributions}
Channel capacity problems often require optimizing the input distribution $p_X$.
Our framework can naturally incorporate 
this optimization by viewing the input generation process itself 
{\em as part of the DAG}.
Any distribution that can be sampled from using a computational procedure (e.g., inverse transform sampling, Gaussian mixture models, 
or deep generative models like VAE, rectified flow \cite{liu2023rectifiedflow}) 
can be represented as a sub-graph 
where the root nodes are primitive noise sources (e.g., multi-dimensional standard Gaussian 
or uniform random variables).
By prepending this generative sub-graph to the main channel DAG, 
the parameters of the input distribution become standard network parameters $\bm \eta$.
This enables unified end-to-end gradient-based optimization of both the source distribution 
and the channel/system parameters simultaneously, 
offering a powerful numerical approach for bounding the capacity region of the channel.

\subsection{Unified Procedure for Rate Region Exploration}

The information gradient framework developed in this paper provides 
a unified numerical procedure for exploring rate regions of multi-user 
channels. For a given parameterization of the input distributions 
and system components, the resulting rate region 
defined by mutual information expressions constitutes 
a  bound on the capacity region. 
When the parameterization is sufficiently expressive to represent 
the capacity-achieving input distributions, this  bound 
numerically coincides 
with the true capacity region. We outline the general methodology, 
which applies to any channel representable as a DAG, 
regardless of whether analytical solutions exist.

\subsubsection{Weighted Sum Maximization}

The boundary of a rate region can be characterized as 
the Pareto frontier of a multi-objective optimization problem 
involving multiple mutual information expressions. 
A standard approach to trace this boundary is to solve a family of 
weighted optimization problems parameterized 
by $\bm{\lambda} = (\lambda_1, \ldots, \lambda_K)$ with $\lambda_i \geq 0$:
\begin{equation}
    \max_{\bm{\theta}} \sum_{i=1}^{K} \lambda_i R_i(\bm{\theta}) \quad \text{subject to resource constraints},
    \label{eq:weighted_rate}
\end{equation}
where $R_i(\bm{\theta})$ denotes the $i$-th mutual information expression, 
which may be an unconditional mutual information $I(X_i; Y)$ or 
a conditional mutual information $I(X_i; Y | X_j)$, and $\bm{\theta}$ 
represents the tunable parameters of the system 
(e.g., input distribution parameters, encoder weights, or power allocations). 
By sweeping $\bm{\lambda}$ over the probability simplex 
and solving \eqref{eq:weighted_rate} for each weight configuration, 
one obtains a collection of boundary points that trace out a bound 
on the rate region.

\subsubsection{Information Gradient for Conditional Mutual Information}

To apply gradient-based optimization to \eqref{eq:weighted_rate}, we require the gradient of conditional mutual information with respect to system parameters. Theorem~\ref{thm:dag_gradient} extends naturally to conditional mutual information as follows.

\begin{corollary}[Information Gradient for Conditional MI]
\label{cor:conditional_mi}
For a DAG channel with output $Y$ depending on inputs $X_1, X_2$ and parameter $\bm{\theta}$, the gradient of the conditional mutual information $I(X_1; Y | X_2)$ is given by
\begin{align} \nonumber 
    &\nabla_{\bm{\theta}} I(X_1; Y | X_2)  \\
    &= \mathbb{E}\left[ (D_{\bm{\theta}} Y)^\top \left( s_{Y|X_1, X_2}(Y|X_1, X_2) - s_{Y|X_2}(Y|X_2) \right) \right],
    \label{eq:cond_mi_gradient}
\end{align}
where $s_{Y|X_1,X_2}$ and $s_{Y|X_2}$ are the conditional score functions.
\end{corollary}

\begin{IEEEproof}
This follows directly from the decomposition $I(X_1; Y | X_2) = h(Y|X_2) - h(Y|X_1, X_2)$ and 
applying the same derivation technique as 
in Theorem~\ref{thm:dag_gradient} to each conditional 
entropy term.
\end{IEEEproof}

This result, combined with Theorem~\ref{thm:dag_gradient}, enables the computation 
of gradients for all mutual information expressions appearing in standard multi-user 
channel characterizations, including MAC sum-rate $I(X_1, X_2; Y)$ and individual rates $I(X_i; Y | X_j)$.

\subsubsection{Algorithmic Framework}

For concreteness, we describe the procedure for a 
two-user channel (e.g., MAC) where the rate region is characterized 
by two mutual information expressions $R_1(\bm{\theta})$ and $R_2(\bm{\theta})$. 
Algorithm~\ref{alg:rate_region} outlines the boundary exploration procedure. 
The algorithm sweeps a scalar weight $\lambda \in [0, 1]$ over a uniform grid, 
solves the weighted optimization problem $\max_{\bm{\theta}} 
\lambda R_1 + (1-\lambda) R_2$ for each $\lambda$, and finally computes 
the convex hull of the resulting points to obtain the rate region.

\begin{algorithm}[t]
\caption{Rate Region Boundary Exploration (Two-User Case)}
\label{alg:rate_region}
\begin{algorithmic}[1]
\Require DAG structure $\mathcal{G}$, mutual information expressions $R_1, R_2$, number of weight samples $L$, constraint set $\mathcal{C}$
\Ensure Achievable rate region $\mathcal{R}$
\State $\mathcal{B} \gets \emptyset$ \Comment{Set of boundary points}
\For{$l = 0$ to $L$}
    \State $\lambda \gets l / L$ \Comment{Weight sweep}
    \State Initialize parameters $\bm{\theta}$
    \Repeat
        \State Generate samples from DAG $\mathcal{G}$ with current $\bm{\theta}$
        \State Estimate required score functions via DSM
        \State Compute $\nabla_{\bm{\theta}} R_1$ and $\nabla_{\bm{\theta}} R_2$ using \eqref{eq:cond_mi_gradient}
        \State $\nabla_{\bm{\theta}} \mathcal{L} \gets \lambda \nabla_{\bm{\theta}} R_1 + (1 - \lambda) \nabla_{\bm{\theta}} R_2$
        \State $\bm{\theta} \gets \mathcal{P}_{\mathcal{C}}(\bm{\theta} + \alpha \nabla_{\bm{\theta}} \mathcal{L})$ \Comment{Projected gradient ascent}
    \Until{convergence}
    \State Evaluate $(R_1^*, R_2^*) \gets (R_1(\bm{\theta}), R_2(\bm{\theta}))$  (e.g., via Fisher integral 
     or analytical expressions if available)
    \State $\mathcal{B} \gets \mathcal{B} \cup \{(R_1^*, R_2^*)\}$
\EndFor
\State $\mathcal{R} \gets \mathrm{ConvexHull}(\mathcal{B})$ \Comment{Time-sharing achievability}
\Return $\mathcal{R}$
\end{algorithmic}
\end{algorithm}

The final convex hull operation reflects the fact that any convex combination 
of achievable rate pairs is also achievable via time-sharing 
between the corresponding coding strategies.

A key practical consideration is \emph{score drift}: 
as $\bm{\theta}$ evolves during optimization, 
the underlying distributions change, requiring 
the score models to be updated. 
This can be addressed either by periodic retraining 
of the score networks or by employing conditional score 
models $s_\phi(y; \bm{\theta})$ that take the system 
parameters as additional inputs, as discussed 
in Section~\ref{sec:score_drift}.

\subsubsection{Scope and Future Directions}

The framework presented here establishes the theoretical and algorithmic foundations 
for rate region exploration via information gradients. While the linear Gaussian case 
admits analytical verification (see Section~\ref{sec:gaussian_mac_validation}), the primary value 
of this approach lies in its applicability to channels where closed-form solutions are unavailable, 
such as nonlinear MACs with $Y = f(X_1, X_2) + Z$ or channels with non-Gaussian noise. 
For such channels, Algorithm~\ref{alg:rate_region} provides a systematic numerical method 
to explore bounds on the capacity region. Comprehensive numerical studies for specific 
nonlinear multi-user channels constitute a promising direction for future research.

\subsection{Analytical Validation: Gaussian MAC}
\label{sec:gaussian_mac_validation}

To validate Algorithm~\ref{alg:rate_region}, we consider the linear Gaussian MAC 
with a sum power constraint. We derive the rate region via two independent routes 
and demonstrate their equivalence: Route~1 uses the information gradient formula 
(Corollary~\ref{cor:conditional_mi}) with analytical score functions, 
while Route~2 uses direct differentiation of the known mutual information expressions. 
The agreement of the resulting rate regions confirms the correctness of the proposed framework.

\subsubsection{Channel Model}

Consider a two-user Gaussian MAC defined by
\begin{equation}
    Y = X_1 + X_2 + Z,
    \label{eq:gaussian_mac}
\end{equation}
where $X_1 \sim \mathcal{N}(0, P_1)$ and $X_2 \sim \mathcal{N}(0, P_2)$ are independent Gaussian inputs, 
$Z \sim \mathcal{N}(0, \sigma^2)$ is independent Gaussian noise, 
and we impose a sum power constraint $P_1 + P_2 \leq P$. The rate expressions are:
\begin{align}
    R_1(P_1) &= I(X_1; Y | X_2) = \frac{1}{2} \log \left( 1 + \frac{P_1}{\sigma^2} \right), \label{eq:mac_r1} \\
    R_2(P_2) &= I(X_2; Y | X_1) = \frac{1}{2} \log \left( 1 + \frac{P_2}{\sigma^2} \right). \label{eq:mac_r2}
\end{align}

\subsubsection{Route 1: Score-Based Derivation}

We compute the rate region using the information gradient framework. 
The analytical score functions for this Gaussian channel are:
\begin{align}
    s_{Y|X_1,X_2}(y|x_1,x_2) &= -\frac{y - x_1 - x_2}{\sigma^2}, \\
    s_{Y|X_2}(y|x_2) &= -\frac{y - x_2}{P_1 + \sigma^2}.
\end{align}
Using the parameterization $X_1 = \sqrt{P_1}\tilde{X}_1$ with $\tilde{X}_1 \sim \mathcal{N}(0,1)$, we have $D_{P_1}Y = X_1/(2P_1)$. Applying Corollary~\ref{cor:conditional_mi}:
\begin{equation}
    \nabla_{P_1} R_1 = \mathbb{E}\left[ \frac{X_1}{2P_1} \left( s_{Y|X_1,X_2} - s_{Y|X_2} \right) \right] = \frac{1}{2(P_1 + \sigma^2)}.
    \label{eq:grad_route1_r1}
\end{equation}
By symmetry, $\nabla_{P_2} R_2 = 1/(2(P_2 + \sigma^2))$.

For the weighted objective $\mathcal{L} = \lambda R_1 + (1-\lambda)R_2$ subject to $P_1 + P_2 = P$, the KKT stationarity conditions using these gradients are:
\begin{equation}
    \frac{\lambda}{2(P_1 + \sigma^2)} = \mu, \quad \frac{1-\lambda}{2(P_2 + \sigma^2)} = \mu.
    \label{eq:kkt_route1}
\end{equation}
Eliminating $\mu$ and solving with $P_1 + P_2 = P$:
\begin{align}
    P_1^{(1)}(\lambda) &= \lambda(P + 2\sigma^2) - \sigma^2, \label{eq:p1_route1} \\
    P_2^{(1)}(\lambda) &= (1-\lambda)(P + 2\sigma^2) - \sigma^2. \label{eq:p2_route1}
\end{align}
Substituting into the rate expressions, the boundary points are:
\begin{align}
    R_1^{(1)}(\lambda) &= \frac{1}{2}\log\left( \frac{\lambda(P + 2\sigma^2)}{\sigma^2} \right), \label{eq:r1_route1} \\
    R_2^{(1)}(\lambda) &= \frac{1}{2}\log\left( \frac{(1-\lambda)(P + 2\sigma^2)}{\sigma^2} \right). \label{eq:r2_route1}
\end{align}

\subsubsection{Route 2: Classical Derivation}

We compute the rate region by directly differentiating the known expressions \eqref{eq:mac_r1}--\eqref{eq:mac_r2}:
\begin{equation}
    \nabla_{P_1} R_1 = \frac{1}{2(P_1 + \sigma^2)}, \quad \nabla_{P_2} R_2 = \frac{1}{2(P_2 + \sigma^2)}.
    \label{eq:grad_route2}
\end{equation}
The KKT conditions for $\mathcal{L} = \lambda R_1 + (1-\lambda)R_2$ subject to $P_1 + P_2 = P$ are identical to \eqref{eq:kkt_route1}. Solving yields:
\begin{align}
    P_1^{(2)}(\lambda) &= \lambda(P + 2\sigma^2) - \sigma^2, \label{eq:p1_route2} \\
    P_2^{(2)}(\lambda) &= (1-\lambda)(P + 2\sigma^2) - \sigma^2. \label{eq:p2_route2}
\end{align}
The corresponding boundary points are:
\begin{align}
    R_1^{(2)}(\lambda) &= \frac{1}{2}\log\left( \frac{\lambda(P + 2\sigma^2)}{\sigma^2} \right), \label{eq:r1_route2} \\
    R_2^{(2)}(\lambda) &= \frac{1}{2}\log\left( \frac{(1-\lambda)(P + 2\sigma^2)}{\sigma^2} \right). \label{eq:r2_route2}
\end{align}

\subsubsection{Equivalence of Rate Regions}

Comparing \eqref{eq:r1_route1}--\eqref{eq:r2_route1} with \eqref{eq:r1_route2}--\eqref{eq:r2_route2}, we confirm:
\begin{equation}
    \left( R_1^{(1)}(\lambda), R_2^{(1)}(\lambda) \right) = \left( R_1^{(2)}(\lambda), R_2^{(2)}(\lambda) \right), \quad \forall \lambda \in [0, 1].
    \label{eq:region_equivalence}
\end{equation}
The two routes produce identical rate region boundaries. 
Including the corner points at $\lambda \in \{0, 1\}$ where $(R_1, R_2) = (0, \frac{1}{2}\log(1+P/\sigma^2))$ or $(\frac{1}{2}\log(1+P/\sigma^2), 0)$, and taking the convex hull, both routes yield the same achievable rate region.

\subsubsection{Implications}

This analytical validation demonstrates that Algorithm~\ref{alg:rate_region}, which implements 
Route~1 computationally, exactly reproduces the known capacity region derived via Route~2. 
The key insight is that the score-based information gradient (Corollary~\ref{cor:conditional_mi}) 
provides the correct optimization direction, leading to the same optimal power allocation 
and hence the same rate region. For nonlinear or non-Gaussian channels where Route~2 is 
unavailable---since closed-form mutual information expressions do not exist---Route~1 with 
learned score functions provides the only systematic approach to explore rate region boundaries.

\section{Numerical Experiments}

\subsection{Information Gradient Estimation on Gaussian Multipath DAG}

Before moving on to more complex nonlinear and high-dimensional examples, we 
first provide a numerical sanity check of 
Theorem~\ref{thm:dag_gradient} on the linear Gaussian multipath DAG 
introduced in Sec.~\ref{subsec:multipath-dag}.
In that subsection, the DAG structure
\[
  X \to (V_1, V_2) \to Y
\]
was specified, and the corresponding effective scalar Gaussian channel
$Y = G_{\mathrm{eff}} X + Z_{\mathrm{eff}}$ was derived, together with the
closed-form expressions of the mutual information $I(X;Y)$ and its gradients 
with respect to $(\eta_1,\eta_2)$.
In particular, the analytic gradients
\[
  \frac{\partial I}{\partial \eta_1}, \quad
  \frac{\partial I}{\partial \eta_2}
\]
were obtained in \eqref{eq:true_grad_eta1} and \eqref{eq:true_grad_eta2},
and their equality with the DAG gradient formula in
Theorem~\ref{thm:dag_gradient} was verified analytically using the
Gaussian score functions and the Jacobians $D_{\eta_k}Y$.
In the present subsection, we complement this analysis by a fully
data-driven Monte Carlo study of the same model.

\subsubsection{Experimental setup}

We reuse exactly the same linear Gaussian multipath DAG as in Sec.~IV-C,
including the variance parameters and the definitions of $G_{\mathrm{eff}}$,
$\sigma_N^2$, and $v_Y$.
Unless otherwise stated, we adopt the normalized setting
\begin{equation}
  \sigma_X^2 = \sigma_1^2 = \sigma_2^2 = \sigma_3^2 = 1,
\end{equation}
so that the input and all noise branches have unit variance.
We consider two one-dimensional sweeps of the parameters:
\begin{itemize}
  \item For the first sweep, we fix $\eta_2 = \eta_{2,\mathrm{fix}} = 1$
        and vary $\eta_1$ over $[\eta_{1,\min},\eta_{1,\max}] = [-2,2]$
        using $K_1$ equispaced grid points.
  \item For the second sweep, we fix $\eta_1 = \eta_{1,\mathrm{fix}} = 1$
        and vary $\eta_2$ over $[\eta_{2,\min},\eta_{2,\max}] = [-2,2]$
        with $K_2$ equispaced grid points.
\end{itemize}
For each grid point $(\eta_1,\eta_2)$ we evaluate the gradients
$\partial I/\partial\eta_1$ and $\partial I/\partial\eta_2$ in three
different ways:
\begin{enumerate}
  \item[(i)] \emph{Analytic gradient:} We directly evaluate the closed-form
    expressions \eqref{eq:true_grad_eta1} and \eqref{eq:true_grad_eta2}
    derived in Sec.~IV-C; these are treated as ground truth.
  \item[(ii)] \emph{Monte Carlo gradient with analytic scores:}
    We apply Theorem~\ref{thm:dag_gradient} to the multipath DAG, using
    the same Gaussian score functions $s_Y(y)$ and $s_{Y|X}(y|x)$ as in
    Sec.~IV-C and computing the Jacobians $D_{\eta_k}Y$ via reverse-mode
    automatic differentiation (VJP).
    Concretely, we estimate
    \begin{equation}
      \frac{\partial I}{\partial \eta_k}
        \approx \frac{1}{N}
        \sum_{n=1}^N
        \big( D_{\eta_k} Y^{(n)} \big)
        \big( s_{Y|X}(Y^{(n)}|X^{(n)}) - s_Y(Y^{(n)}) \big),
    \end{equation}
    where $(X^{(n)},Y^{(n)})$ are i.i.d.\ samples generated by the DAG.
  \item[(iii)] \emph{Monte Carlo gradient with fully learned scores:}
    We again use Theorem~\ref{thm:dag_gradient}, but now replace {both} the
    unconditional score $s_Y(y)$ {and the conditional score $s_{Y|X}(y|x)$}
    by neural score estimators {$\hat{s}_Y(y)$ and $\hat{s}_{Y|X}(y|x)$}
    obtained by denoising score matching (DSM).
    This setup assumes no knowledge of the true densities, validating the
    method as a black-box estimator.
\end{enumerate}

For the Monte Carlo estimates (ii) and (iii), we use $N = 10^5$ i.i.d.\ samples
$(X^{(n)}, Z_1^{(n)}, Z_2^{(n)}, Z_3^{(n)})$, generate the corresponding
$Y^{(n)}$ through the multipath DAG, and evaluate the gradients with a
mini-batch size of $8{,}192$.

\subsubsection{Score learning details}

For each grid point $(\eta_1,\eta_2)$ we learn score models specific to that
channel realization.
The training data consists of clean samples sampled from the multipath
DAG in Sec.~IV-C.
{We employ two separate MLPs: one for the unconditional score $\hat{s}_Y(y)$
and another for the conditional score $\hat{s}_{Y|X}(y|x)$ (which takes the
concatenation of $y$ and $x$ as input).}
Each network has three hidden layers of width $128$, SiLU activations, and a
linear output layer.

We train $\hat{s}_Y$ and \textbf{$\hat{s}_{Y|X}$} by denoising score matching
at a fixed noise variance $t_{\mathrm{DSM}} = 0.05$.
The unconditional score is trained to denoise $Y$, while the conditional score
is trained to denoise $Y$ given $X$.
In the reported experiments we use a batch size of $4{,}096$, $500$ gradient
steps, and the AdamW optimizer with learning rate $10^{-3}$.

After training, we apply a scalar Stein calibration to correct global
scaling bias of the learned {unconditional} score $\hat{s}_Y$.
For the one-dimensional output this amounts to the rescaling
\begin{equation}
  \tilde{s}_Y(y) = c \,\hat{s}_Y(y),
  \qquad
  c = - \frac{1}{\mathbb{E}[Y \hat{s}_Y(Y)]},
\end{equation}
which enforces $\mathbb{E}[Y \tilde{s}_Y(Y)] = -1$.
In the Monte Carlo gradient (iii), we use $\tilde{s}_Y$ in place of
$\hat{s}_Y$ {(combined with the raw $\hat{s}_{Y|X}$)}.

\subsubsection{Results}

\begin{figure}[t]
  \centering
  \includegraphics[width=1.0\linewidth]{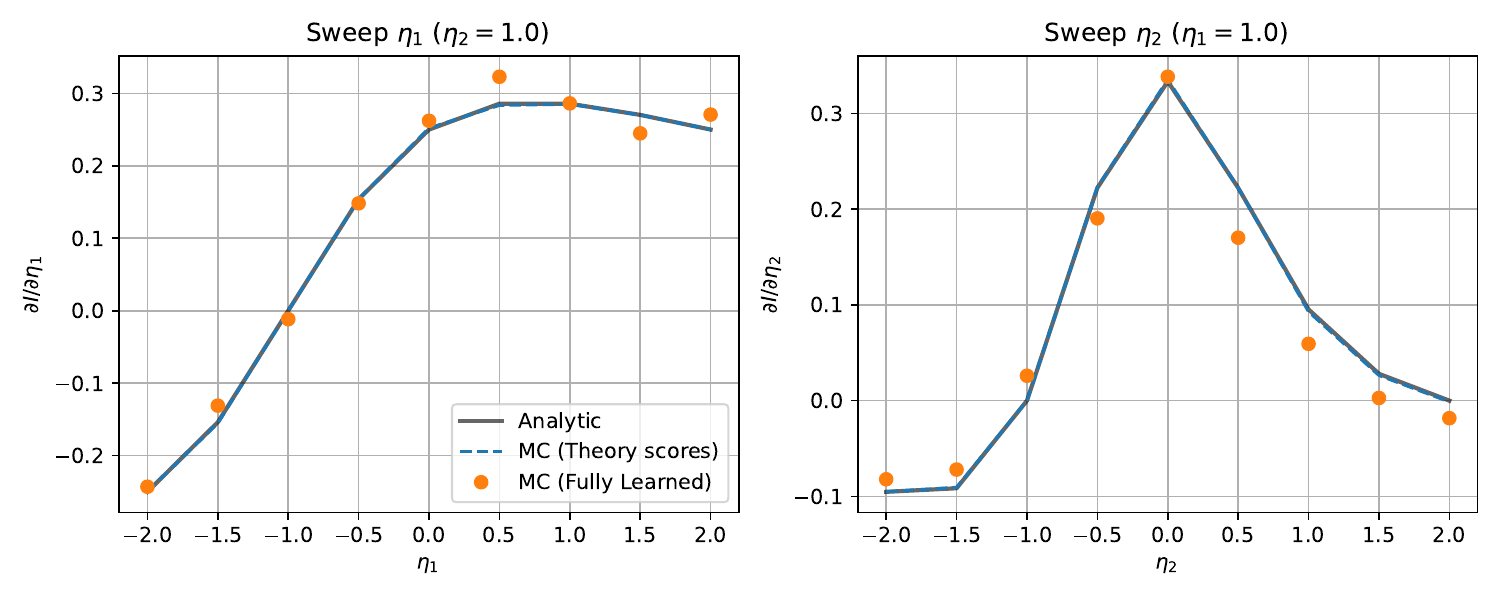}
  \caption{Information gradients for the linear Gaussian multipath DAG from
  Sec.~IV-C. Left: $\partial I/\partial\eta_1$ as a function of $\eta_1$
  with $\eta_2$ fixed. Right: $\partial I/\partial\eta_2$ as a function
  of $\eta_2$ with $\eta_1$ fixed. In each panel, we compare the analytic
  gradient (solid line), the Monte Carlo evaluation of
  Theorem~\ref{thm:dag_gradient} with analytic score functions
  (dashed line), and the Monte Carlo evaluation of
  Theorem~\ref{thm:dag_gradient} with {fully learned scores} (markers).
  }
  \label{fig:multipath-grad}
\end{figure}

Figure~\ref{fig:multipath-grad} summarizes the results.
The left panel shows $\partial I/\partial\eta_1$ as a function of $\eta_1$
with $\eta_2$ fixed, while the right panel shows $\partial I/\partial\eta_2$
as a function of $\eta_2$ with $\eta_1$ fixed.
In each panel we plot:
\begin{itemize}
  \item the analytic gradient
    \eqref{eq:true_grad_eta1} or \eqref{eq:true_grad_eta2}
    (solid line),
  \item the Monte Carlo estimate based on
    Theorem~\ref{thm:dag_gradient} with analytic
    scores $s_{Y|X}$ and $s_Y$ (dashed line), and
  \item the Monte Carlo estimate based on
    Theorem~\ref{thm:dag_gradient} with {fully learned scores
    $\hat{s}_{Y|X}$ and $\tilde{s}_Y$} (markers).
\end{itemize}

The analytic gradient and the Monte Carlo gradient with analytic scores are
indistinguishable at the scale of the plots for both $\eta_1$- and
$\eta_2$-sweeps, providing a strong sanity check of the gradient formula in
Theorem~\ref{thm:dag_gradient} as well as its VJP-based implementation.
Furthermore, the data-driven gradients obtained from the DSM-trained {score
networks} closely track the analytic curves across the entire range of
$(\eta_1,\eta_2)$.
This demonstrates that the proposed score-learning approach can faithfully
reproduce the information gradient even when the DAG is treated as a
{fully} black-box generator of $(X,Y)$ samples, and complements the analytic
verification given in Sec.~\ref{subsec:multipath-dag}.

\subsection{Optimization of Multipath DAG under Global Constraint}

We apply the proposed information maximization framework to the linear Gaussian
multipath DAG defined in Sec.~\ref{subsec:multipath-dag}.
The goal is to maximize the end-to-end mutual information $I(X;Y)$ with
respect to the parameters $\eta = (\eta_1, \eta_2)$ subject to a global
power constraint $\|\eta\|_2^2 = 1$.
This problem serves as a dynamic validation of our framework, assessing whether
the gradient estimator remains robust as the underlying data distribution shifts
during the optimization process.

\subsubsection{Experimental Setup}

We initialize the parameters at $(\eta_1, \eta_2) = (1.0, 0.0)$.
We employ the PGA algorithm with a fixed step size
of $\alpha = 0.1$ for $35$ iterations.
At each iteration, we estimate the gradient $\nabla_\eta I(X;Y)$ using
Theorem~\ref{thm:dag_gradient}.
Crucially, to test the fully black-box capability, we assume no knowledge of the
analytic score functions.
Instead, we estimate the gradient using {fully learned scores}, where
both the unconditional score $s_Y(y)$ and the conditional score $s_{Y|X}(y|x)$
are approximated by neural networks.

\subsubsection{Iterative Score Learning Strategy}

Since the distribution of $Y$ (and $Y|X$) changes as $\bm \eta$ evolves, the
score networks must be updated at each PGA step.
To handle this efficiently, we adopt a \emph{warm-start} strategy.
We maintain two score networks, $\hat{s}_Y$ and $\hat{s}_{Y|X}$, which persist
across iterations.
At the first iteration ($t=0$), these networks are trained from scratch
for $1{,}000$ steps using DSM to ensure accurate initial gradients.
In subsequent iterations ($t > 0$), we update the networks using only $200$
steps of DSM, initializing weights from the previous iteration.
This fine-tuning approach significantly reduces computational cost while
allowing the score models to track the drifting distributions effectively.

Both networks are MLPs with the same architecture as in the previous subsection.
For calibration, we apply Stein calibration to {both} the unconditional
and conditional scores at every iteration.
Specifically, for the conditional score $\hat{s}_{Y|X}$, we compute a scalar
correction factor $c$ satisfying
$\mathbb{E}_{X}[\mathbb{E}_{Y|X}[\langle Y - \mathbb{E}[Y|X], c\,\hat{s}_{Y|X}(Y|X) \rangle]] = -1$,
ensuring robust scale estimates throughout the optimization trajectory.

\subsubsection{Results}

\begin{figure}[t]
  \centering
  \includegraphics[width=1.0\linewidth]{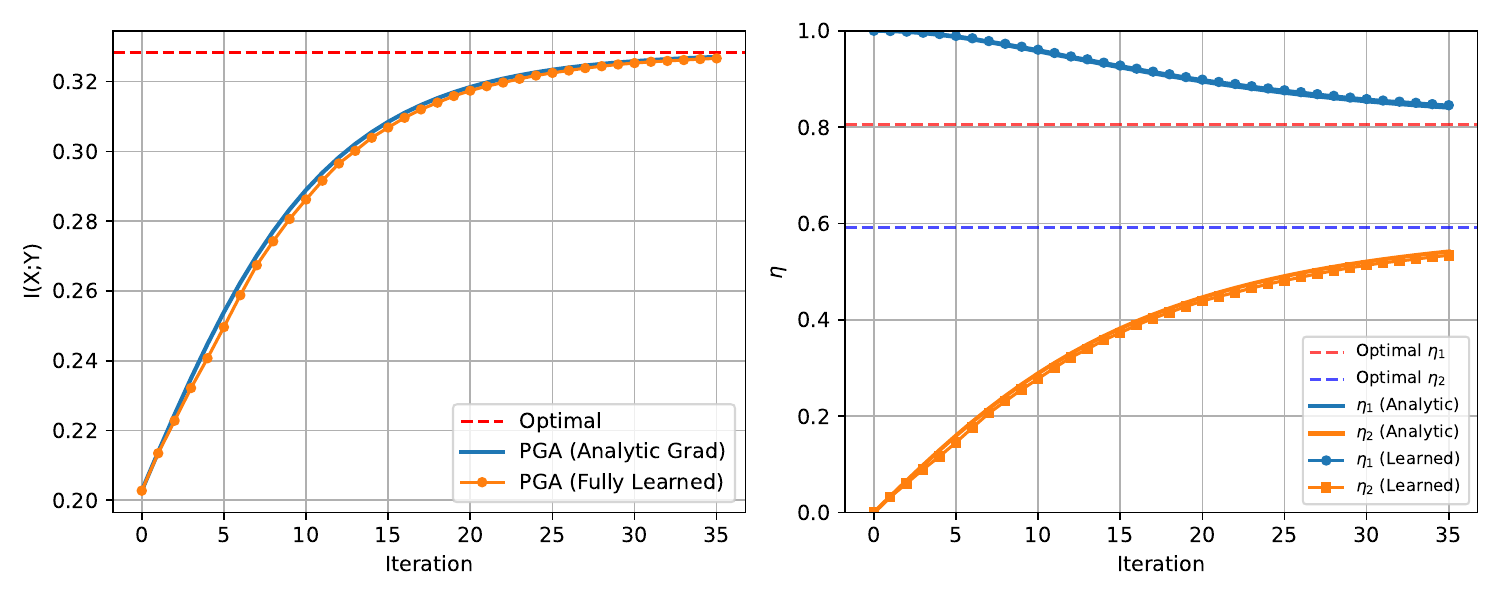}
  \caption{Maximization of mutual information for the multipath DAG under
  the constraint $\eta_1^2 + \eta_2^2 = 1$.
  Left: Evolution of $I(X;Y)$ over PGA iterations.
  Right: Trajectories of parameters $\eta_1$ and $\eta_2$.
  The proposed fully learned method (markers), which learns both $s_Y$ and
  $s_{Y|X}$ online with iterative fine-tuning and dual Stein calibration,
  perfectly tracks the optimal trajectory computed via analytic gradients
  (solid lines).}
  \label{fig:pga-traj}
\end{figure}

Figure~\ref{fig:pga-traj} presents the optimization results.
The left panel shows the trajectory of the mutual information $I(X;Y)$ over
iterations, comparing the PGA using the analytic gradient (solid line) and the
PGA using the fully learned gradient (markers).
The right panel displays the corresponding trajectories of the parameters
$\eta_1$ and $\eta_2$.

The results demonstrate remarkable agreement between the proposed method and
the ground-truth analytic optimization.
The fully learned approach steadily maximizes the mutual information,
converging to the theoretical optimum (indicated by the red dashed line)
at the same rate as the analytic baseline.
Furthermore, the right panel confirms that the parameters $(\eta_1, \eta_2)$
follow the correct path on the constraint manifold.
This experiment confirms that our framework, equipped with the iterative
fine-tuning strategy, can successfully drive end-to-end optimization in
dynamic settings without access to closed-form likelihoods.

\subsection{Nonlinear scalar channel}

Here, we verify that the proposed information-gradient estimator also works
reliably for a genuinely nonlinear channel where no closed-form mutual
information is available. We consider the scalar model:
\begin{equation}
  X \sim \mathcal{N}(0,1),\quad
  Z \sim \mathcal{N}(0,\sigma_Z^2),\quad
  Y = \tanh(\eta X) + Z,
\end{equation}
with fixed noise variance $\sigma_Z^2 = 0.25$ and scalar parameter
$\eta \in [\eta_{\min},\eta_{\max}] = [-3,3]$.
For each value of $\eta$ on a small grid, we compute
$I(X;Y)$ numerically from the definition $I(X;Y) = h(Y) - h(Z)$.
Here $h(Z)$ is analytic, whereas $h(Y)$ is evaluated by first computing
$p_Y(y) = \int p_X(x)p_{Y|X}(y|x)\,dx$ via high-order Gauss--Hermite
quadrature over $x$, and then approximating
$H(Y) = -\int p_Y(y)\log p_Y(y)\,dy$ on a finite interval by a
uniform $y$-grid and the trapezoid rule.

The ground-truth gradient $\partial I/\partial\eta$ is approximated from
these values by a centered finite difference
\begin{equation}
  \frac{\partial I}{\partial\eta}(\eta)
  \approx \frac{I(\eta+h) - I(\eta-h)}{2h},
\end{equation}
with a small stepsize $h = 0.01$.
In parallel, we estimate the information gradient using
Theorem~\ref{thm:dag_gradient} with the   {fully learned scores} implementation:
{both} the unconditional score $s_Y(y)$ {and the conditional score
$s_{Y|X}(y|x)$ are approximated by DSM-trained score networks.}
Furthermore, {both scores are corrected via scalar Stein calibration.}
The expectation in Theorem~\ref{thm:dag_gradient} is evaluated by Monte Carlo
using $N = 10^5$ samples.

\begin{figure}[t]
  \centering
  \includegraphics[width=1.0\linewidth]{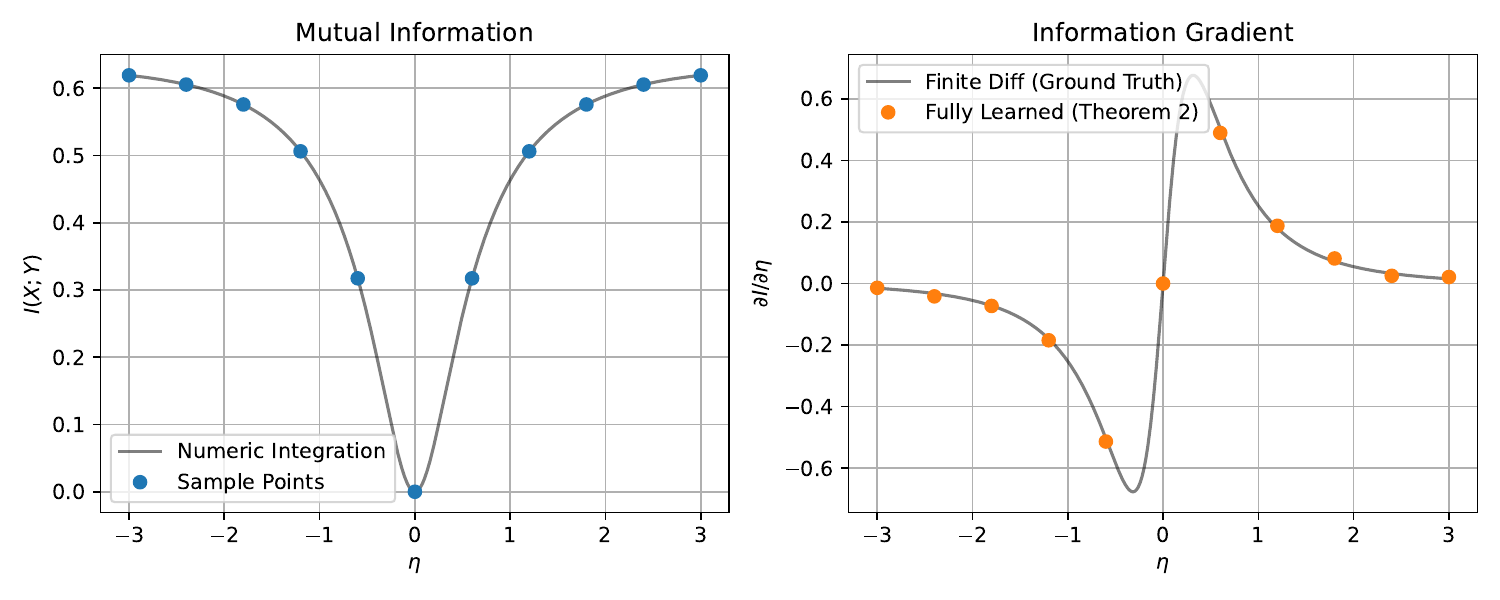}
  \caption{Nonlinear scalar channel $Y = \tanh(\eta X) + Z$ with
  $X \sim \mathcal{N}(0,1)$ and $Z \sim \mathcal{N}(0,0.25)$.
  Left: mutual information $I(X;Y)$ as a function of the nonlinearity
  parameter $\eta$, computed by numerical integration of
  $I(X;Y) = h(Y) -h(Z)$ using Gauss--Hermite quadrature 
  for $p_Y(y)$ and a  uniform $y$-grid.
  Right: information gradient $\partial I/\partial\eta$ versus $\eta$.
  We compare the finite-difference approximation based on the numerically
  evaluated $I(\eta)$ (solid line) with the gradient obtained from
  Theorem~\ref{thm:dag_gradient} using {fully learned and
  Stein-calibrated scores} (markers).}
  \label{fig:tanh-mi-grad}
\end{figure}

Figure~\ref{fig:tanh-mi-grad} summarizes the results.%
\footnote{See also the numerical curves in the supplementary material.}
The left panel shows the mutual information $I(X;Y)$ as a function of $\eta$,
exhibiting the expected even symmetry and saturation as $\lvert\eta\rvert$
increases. The right panel compares the finite-difference gradient and the
information gradient obtained from Theorem~\ref{thm:dag_gradient} with
{fully} learned scores. The two curves are nearly indistinguishable over the entire
range of $\eta$, including around $\eta=0$ where the true gradient vanishes
and in the high-gain regime where $I(X;Y)$ saturates. This experiment
indicates that the proposed score-based information-gradient estimator
remains accurate and stable even for nonlinear channels where $I(X;Y)$
itself is only available via numerical integration.

\section{Extension: Digital Twin Calibration}

The information maximization framework presented so far relies on the premise that 
the DAG model $\mathcal{G}$ accurately represents the real physical system.
If there is a significant discrepancy between the model and reality, 
the parameters $\bm \eta^*$ optimized in the simulation may fail to perform 
as expected in the real world.
Therefore, calibrating the DAG model to match reality, referred to as \textit{digital twin calibration}, 
is a prerequisite for practical deployment.
In this section, we show that our score-based framework can be naturally extended 
to address this calibration problem.

\subsection{Problem Formulation}
Let the real physical system be governed by an unknown stochastic process 
that produces output $Y_{\text{real}}$ from input $X$ and 
internal noise sources $\mathcal{Z}_{\text{real}}$:
\begin{equation}
Y_{\text{real}} = g_{\text{real}}(X, \mathcal{Z}_{\text{real}}).
\end{equation}
Our digital twin is the DAG model defined in \eqref{eq:dag_mapping}:
\begin{equation}
Y_{\bm \eta} = g(X, \mathcal{Z}; \bm \eta).
\end{equation}
We assume that the distributions of inputs $X$ and noise sources $\mathcal{Z}$ 
in the model match those in the real system.
The goal of calibration is to find the parameters $\bm \eta^*$ that minimize 
the discrepancy between the real system $g_{\text{real}}$ and the DAG model $g(\cdot; \bm \eta^*)$.

\subsection{Score-Based Calibration Strategy}

\begin{figure}[htbp]
  \centering
  \includegraphics[width=\linewidth]{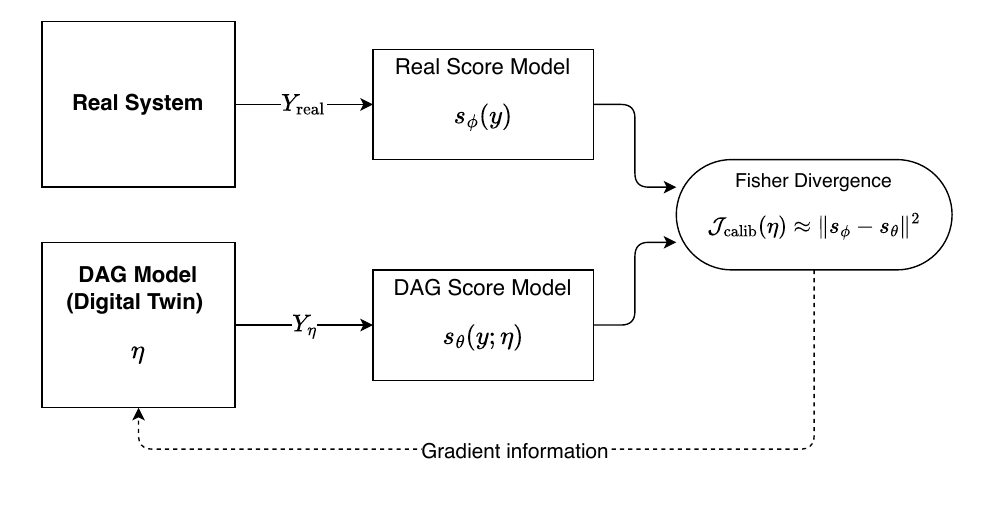}
  \caption{The score-based calibration strategy. The "Real Score Model" ($s_{\bm \phi}$) is trained on observational data $Y_{\text{real}}$ from the physical system. The "DAG Score Model" ($s_{\bm \theta}$) is trained on simulation data $Y_{\bm \eta}$ from the digital twin. The calibration objective $\mathcal{J}_{\text{calib}}(\eta)$ is the Fisher divergence (approximated by the squared difference) between these two score functions. The gradient of this objective is used to update the DAG model parameters $\bm \eta$ to match reality.}
  \label{fig:calibration_strategy}
\end{figure}

Since the true mechanism $g_{\text{real}}$ is unknown and we can only observe samples 
from $Y_{\text{real}}$, we cannot directly compare the internal functions.
Instead, we aim to match their output distributions.
Based on the SFB philosophy, we use the Fisher divergence between 
the real and modeled output distributions as the calibration objective:
\begin{align}
\mathcal{J}_{\text{calib}}(\bm \eta) 
&\equiv D_F(p_{Y_{\text{real}}} \| p_{Y_{\bm \eta}}) \nonumber \\
&= \mathbb{E}_{\bm y \sim p_{Y_{\text{real}}}} 
\left[ \| s_{Y_{\text{real}}}(\bm y) - s_{Y_{\bm \eta}}(\bm y; \bm \eta) \|^2 \right].
\label{eq:calib_loss}
\end{align}
Minimizing this objective forces the DAG model to generate outputs 
that are statistically indistinguishable from the real system's outputs (in terms of their score functions).
Figure \ref{fig:calibration_strategy} shows the score-based calibration strategy.

\subsection{Calibration Algorithm via Fisher Divergence Minimization}
The optimization of \eqref{eq:calib_loss} can be performed practically using two types of score models:
\begin{itemize}
    \item \textbf{Real Score Model $s_{\bm \phi}(y)$:} Learned from observational samples 
    $\{\bm y_{\text{real}}^{(i)}\}$ of the physical system using unsupervised techniques like 
    Sliced Score Matching (SSM), as clean $g_{\text{real}}$ is unknown.
    \item \textbf{DAG Score Model $s_{\bm \theta}(\bm y; \bm \eta)$:} A conditional score 
    network pre-trained on simulation data from the DAG $Y_{\bm \eta} = g(X, \mathcal{Z}; \bm \eta)$ 
    using standard DSM.
\end{itemize}
The calibration parameters $\bm \eta$ are then updated via gradient descent on the estimated Fisher divergence:
\begin{equation}
\bm \eta_{k+1} \leftarrow \bm \eta_k - \alpha \nabla_{\bm \eta} 
\mathbb{E}_{\bm y \sim p_{Y_{\text{real}}}} 
\left[ \| s_{\bm \phi}(\bm y) - s_{\bm \theta}(\bm y; \bm \eta_k) \|^2 \right]. \label{eq:calib_update}
\end{equation}
This approach allows for calibrating complex, nonlinear DAG models purely 
from output observations, leveraging the same score-based toolkit developed for information maximization.

\begin{remark}[Unsupervised Nature of Calibration] \rm 
  Traditional calibration methods typically require paired input-output data 
  $\{(\bm x^{(i)}, \bm y_{\text{real}}^{(i)})\}$ to minimize element-wise error metrics 
  such as mean squared error (MSE). 
  Gathering such perfectly synchronized paired data from complex physical systems 
  can be operationally difficult or expensive.
  A distinct advantage of our score-based approach is its \textit{unsupervised} nature; 
  it only requires unmatched observational samples $\{\bm y_{\text{real}}^{(i)}\}$ 
  from the real system (assuming the input distribution $p_X$ is shared).
  This flexibility significantly broadens the applicability of the framework 
  in real-world scenarios where only output monitoring is feasible.
\end{remark}

\section{Conclusion}

In this paper, we have established a general framework for end-to-end mutual
information maximization in complex stochastic systems represented by
DAGs.
The core contribution is the derivation of the generalized information gradient
formula (Theorem \ref{thm:dag_gradient}), which unifies and extends the results in \cite{wadayama2025-b} to arbitrary
network topologies involving branching, merging, and nonlinear transformations.
We demonstrated that this gradient can be efficiently computed by combining
modern automatic differentiation tools, specifically VJP, 
with score matching techniques.

A pivotal advantage of our framework is its capability to operate in a fully
black-box manner.
As verified in our numerical experiments on both multipath linear DAGs and
nonlinear scalar channels, the proposed estimator utilizing neural score
models---trained via DSMs and refined by Stein
calibration---can faithfully reproduce the true information gradients without
requiring knowledge of the underlying likelihood functions.
The successful optimization of network parameters under global constraints
further confirms the robustness of the proposed iterative score learning
strategy against distribution drifts.

Beyond maximization, we showed that the score-based perspective naturally
extends to unsupervised model calibration.
The proposed digital twin calibration method minimizes the Fisher divergence
between a physical system and its model using only output samples, offering
a practical solution for aligning simulation environments with reality.

While the accuracy of the gradient estimation naturally depends 
on the fidelity of the learned score models, 
the derived theorem itself provides a rigorous and 
universal interface connecting mutual information 
to the score function. As score estimation techniques 
continue to advance, the applicability and precision of this 
framework will naturally expand.

By bridging the gap between fundamental information-theoretic identities and
scalable deep learning computation, this work provides a versatile toolkit
for designing next-generation information systems, ranging from semantic
communications to distributed sensor networks, where optimizing information
flow through complex, unmodeled channels is paramount.

Future work includes applying this framework to rate-distortion analysis 
and distributed source coding.
While this paper primarily focused on maximizing mutual information 
for channel coding, the proposed information gradient framework is inherently 
dualistic and applies equally to minimization problems. A promising direction for future work 
is the application of this framework to rate-distortion analysis 
and distributed source coding. 
By interpreting the DAG as a test channel (encoder) and minimizing the mutual information under 
distortion constraints, our method can serve as a data-driven solver for characterizing the 
achievable rate regions of complex sensor networks, complementary to the channel capacity problems 
discussed herein. 

\section*{Acknowledgments}
This work was supported by JST, CRONOS, Japan Grant Number JPMJCS25N5.

% References

% Placeholder references (replace with actual .bib file)

\end{document}